%% file: paper.tex
\documentclass{vldb}
\pdfoutput=1

\usepackage[english]{babel}
\usepackage{blindtext}
\usepackage[T1]{fontenc}

\usepackage[show]{chato-notes}
\usepackage{cite}
\usepackage{color}
\usepackage{graphicx}
\usepackage{xspace}
\usepackage{algorithmic}
\algsetup{linenosize=\small}
\usepackage[ruled,lined,linesnumbered]{algorithm2e}

\SetCommentSty{mycommfont}
\usepackage{epsfig}
\usepackage{epstopdf}
\usepackage{caption}
\usepackage{xcolor}
\usepackage{balance}
\usepackage{enumitem}
\usepackage{framed}

\usepackage{hyperref}
\hypersetup{
	colorlinks,breaklinks,
	urlcolor=[rgb]{0,0.25,0.75},
	linkcolor=[rgb]{0,0.25,0.75},
	citecolor=[rgb]{0,0.25,0.75}
}

\input{preamble}

\begin{document}

\title{Refutations on ``Debunking the Myths of Influence Maximization: An In-Depth Benchmarking Study''}

\numberofauthors{5} 
\author{
\alignauthor
Wei Lu\\
\affaddr{Rupert Labs}\\
\affaddr{465 Fairchild Drive}\\
\affaddr{Mountain View, CA 94043}\\
\affaddr{w.lu@alumni.ubc.ca}
\alignauthor
Xiaokui Xiao\\
\affaddr{Nanyang Technological Univ.}\\
\affaddr{50 Nanyang Avenue}\\
\affaddr{Singapore 639798}\\
\affaddr{xkxiao@ntu.edu.sg}
\alignauthor
Amit Goyal\\
\affaddr{Google Inc.}\\
\affaddr{1600 Amphitheatre Parkway}\\
\affaddr{Mountain View, CA 94043}\\
\affaddr{goyalamit@google.com}
\and 
\alignauthor
Keke Huang\\
\affaddr{Nanyang Technological Univ.}\\
\affaddr{50 Nanyang Avenue}\\
\affaddr{Singapore 639798}\\
\affaddr{khuang005@ntu.edu.sg}
\alignauthor
Laks V.S. Lakshmanan
\affaddr{Univ. of British Columbia}\\
\affaddr{201-2366 Main Mall}\\
\affaddr{Vancouver, B.C., Canada}\\
\affaddr{laks@cs.ubc.ca}
}

\maketitle

\begin{abstract}
In a recent SIGMOD paper titled {\em ``Debunking the Myths of Influence Maximization: An In-Depth Benchmarking Study''}, Arora et al.\cite{agr17} undertake a performance benchmarking study of several well-known algorithms for influence maximization.
In the process, they contradict several published results, and claim to have unearthed and debunked several ``myths'' that existed around the research of influence maximization. 

It is the goal of this article to examine their claims objectively and critically, and refute the erroneous ones. 
Our investigation discovers that first, the overall experimental methodology in Arora et al.~\cite{agr17} is flawed and leads to scientifically incorrect conclusions. Second, 
the paper %\cite{agr17} 
is riddled with issues specific to a variety of influence maximization algorithms, including buggy experiments, and draws many misleading conclusions regarding those algorithms. 
Importantly, they fail to recognize the trade-off between running time and solution quality, and did not incorporate it correctly in their experimental methodology. 
In this article, we systematically point out the issues present in~\cite{agr17} and refute 11 of their misclaims.
\end{abstract}

\section{Introduction}\label{sec:intro}
\input{laks-intro}

\section{Issues in Experimental Methodology and Settings}\label{sec:ed}
\input{sec-ed}

\section{Refutations on Algorithm-Specific Misclaims}\label{sec:issues}
Not only the paper has an ill-designed experimental methodology as highlighted above, it is also riddled with various misclaims. This section investigates some of the algorithm-specific issues and misclaims.

\subsection{TIM$^+$ vs. IMM}\label{sec:tim-imm}

\input{sec-tim-imm}

\subsection{SimPath vs. LDAG}\label{sec:simpath}
\input{sec-simpath}

\section{Refutations on Other Misclaims}\label{sec:misc}
\input{sec-misc}

\section{Conclusions}\label{sec:concl}
\input{sec-concl}

\subsubsection*{Acknowledgements}
We sincerely thank Yu Yang (SFU) and Jian Pei (SFU) for their feedback on an earlier version of this work, and for contributing the counterexample in Figure~\ref{fig:mu-std-ex} that depicts a flaw in the experimental methodology of Arora et al.~\cite{agr17}.

\bibliographystyle{abbrv}
\bibliography{singlebib}

\appendix
\section{Background on Influence Maximization}\label{sec:background}
\input{sec-prelim}

\section{A Note On CELF vs. CELF++}\label{sec:celfpp}
\input{sec-celfpp}

%\section{{Misrepresentations on Source Code}}
\section{{Importance of Understanding Source Code}} \label{sec:app-unscientific}
\input{sec-appendix-unscientific}

\end{document}

%% file: preamble.tex
\newcommand{\pink}[1]{\textcolor{magenta}{#1}}

\newcommand{\spara}[1]{\smallskip\noindent\textbf{#1.}}

\newcommand{\comment}[1]{}
\newcommand{\eat}[1]{}

\newcommand{\InFullOnly}[1]{}
 
\newtheorem{example}{Example}

\newtheorem{misclaim}{Misclaim}

\def\noflash#1{\setbox0=\hbox{#1}\hbox to 1\wd0{\hfill}}
\newcommand{\A}{\mathcal{A}\xspace}
\newcommand{\B}{\mathcal{B}\xspace}

\newcommand{\rom}[1]{\uppercase\expandafter{\romannumeral #1\relax}}
\newcommand{\lrom}[1]{\lowercase\expandafter{\romannumeral #1\relax}}

\newcommand{\squishlist}{
 \begin{list}{$\bullet$}
  {  \setlength{\itemsep}{0pt}
     \setlength{\parsep}{3pt}
     \setlength{\topsep}{3pt}
     \setlength{\partopsep}{0pt}
     \setlength{\leftmargin}{2em}
     \setlength{\labelwidth}{1.5em}
     \setlength{\labelsep}{0.5em}
} }
\newcommand{\squishlisttight}{
 \begin{list}{$\bullet$}
  { \setlength{\itemsep}{0pt}
    \setlength{\parsep}{0pt}
    \setlength{\topsep}{0pt}
    \setlength{\partopsep}{0pt}
    \setlength{\leftmargin}{2em}
    \setlength{\labelwidth}{1.5em}
    \setlength{\labelsep}{0.5em}
} }

\newcommand{\squishdesc}{
 \begin{list}{}
  {  \setlength{\itemsep}{0pt}
     \setlength{\parsep}{2pt}
     \setlength{\topsep}{2pt}
     \setlength{\partopsep}{0pt}
     \setlength{\leftmargin}{2em}
     \setlength{\labelwidth}{1.5em}
     \setlength{\labelsep}{0.5em}
} }

\newcommand{\squishdesctight}{
 \begin{list}{}
  {  \setlength{\itemsep}{0pt}
     \setlength{\parsep}{0pt}
     \setlength{\topsep}{0pt}
     \setlength{\partopsep}{0pt}
     \setlength{\leftmargin}{1em}
     \setlength{\labelwidth}{1.5em}
     \setlength{\labelsep}{0.5em}
} }

\newcommand{\squishnumlist} {
\newcounter{qcounter}
\begin{list}{\arabic{qcounter}.~}{\usecounter{qcounter}} 
{  \setlength{\itemsep}{0pt}
    \setlength{\parsep}{0pt}
    \setlength{\topsep}{0pt}
    \setlength{\partopsep}{0pt}
    \setlength{\leftmargin}{2em}
    \setlength{\labelwidth}{1.5em}
    \setlength{\labelsep}{0.5em}
}}

\newcommand{\squishend}{
  \end{list}
}

%% file: laks-intro.tex
Influence maximization in social networks is a well-studied problem with applications in viral marketing, the study of propagation of infections and innovation, and community detection, to name a few. Over the last decade and a half, substantial amount of research has been conducted on this problem, resulting in a plethora of algorithms ranging from heuristics to approximation algorithms, with varying levels of quality, running time efficiency, and memory consumption. In a recent paper titled {\it ``Debunking the Myths of Influence Maximization: An In-Depth Benchmarking Study''} (SIGMOD 2017), Arora et al. \cite{agr17} undertake a performance benchmarking study of several well-known algorithms for influence maximization. In the process, they claim to unearth several ``myths'' which they then claim to debunk. In this article, we examine their claims objectively and critically, point out the errors in many of their claims and systematically refute them. We do this by trying to reproduce their claimed experimental results. Importantly, our analysis shows that first, the overall experimental methodology employed by Arora et al. \cite{agr17} is flawed and leads to scientifically incorrect conclusions. Second, a series of experimental results claimed in \cite{agr17} are a direct consequnce of improper data preparation. We directly refute these claims by running those experiments with proper dataset preparation, and by running independent experiments in two different institutions. Our findings squarely contradict their claims. Third, we identify misleading claims made in \cite{agr17} that fail to properly take into account the trade-off between running time and memory consumption on one hand and accuracy and quality of the solution obtained on the other. We illustrate with examples that these misleading claims can be used to draw obviously incorrect conclusions. 

\subsection{Influence Maximization Recap} 

We begin with a quick review of the problem. Influence Maximization (IM) is an optimization problem studied extensively in the social network data mining literature during the past decade and a half.
Originally motivated by viral marketing~\cite{domingos01,richardson02}, IM was first formulated as a discrete optimization problem in a seminal paper by Kempe et al.~\cite{kempe03}. 
We are given (1) a directed graph $G = (V,E,p)$ as a social network, where nodes are individuals, edges represent relationships, and $p : E \to [0, 1]$ specifies pairwise influence probabilities (or weights) between nodes; (2) a positive integer $k$, and (3) a stochastic diffusion model $M$ which specifies probabilistic rules on how influence propagates from one node to another in the graph. In such a network, activating a set of nodes $S\subseteq V$ leads to a cascade of actions in which the nodes $S$, typically called \emph{seed nodes}, activate at time $t=0$, and activation propagates between nodes in discrete time steps according to the model $M$. Two popular diffusion models in the iterature are \emph{independent cascades} (IC) and \emph{linear threshold} (LT). We refer the reader to \cite{kempe03} for their description. Since $M$ is stochastic, the number of nodes that activate in a cascade is a random number. The \emph{expected spread} of $S$ is defined as the expected total number of activated nodes in a random cascade started by seed nodes $S$. Formally, the expected spread is a function $\sigma_M: 2^{V}\to \mathbb{R}$ that maps every set of nodes to a positive real number, representing the expected number of activated nodes over all possible cascades started by seed nodes $S$. 
 
The optimization problem is to find a set $S\subseteq V$ of $k$ seeds, such that by activating $S$ the expected spread of $S$ under model $M$, denoted $\sigma_M(S)$ (or just $\sigma(S)$ when $M$ is understood),  is maximized.
\eat{ 
Expected spread is thus the objective function in IM, canonically denoted by $\sigma: 2^{|V|}\to \mathbb{R}$ (referred to as ``the spread function'' hereafter).
%A stochastic diffusion model $M$ specifies probabilistic rules on how nodes become 
} 
IM is a computationally challenging problem: (i) it is NP-hard under a large family of stochastic diffusion models~\cite{kempe03} and (ii) for a given set of nodes $S$, computing the spread function $\sigma(S)$ is \#P-hard under both IC and LT models~\cite{ChenWW10,ChenWW10b}.
In response to this two-fold challenge, there has been a large body of research on efficient and scalable IM algorithms, e.g., \cite{leskovec07, KimuraS06, ChenWW10, ChenWW10b, celfpp, simpath, tang14, tang15, borgs14, ChenWY09, ssa16, irie, cohen14}. For convenience, 
technical descriptions of the algorithms relevant to this article are provided in Appendix~\ref{sec:background}.

In their paper, Arora et al. \cite{agr17} undertake a performance benchmarking study of several well-known algorithms for influence maximization, including \cite{leskovec07, ChenWW10, ChenWW10b, celfpp, simpath, tang14, tang15, irie}.
%Unfortunately, \cite{agr17} is riddled with flaws and mis-claims.
In this work, we critically examine the experimental methodology used in \cite{agr17} and their various claims against the IM algorithms proposed in the papers cited above.

\subsection{Overview of Flaws and Misclaims in Arora et al.~\cite{agr17}}

\spara{Flaws in Experimental Design}
One of the most emphasized aspects in Arora et al.~\cite{agr17} is the running time comparison.
To this end, they focus on the following question (\cite{agr17}, Section 5.1.1): 
%{\em ``What is the time taken by a technique to provide its near-optimal quality?''} (see Section 5.1.1 of \cite{agr17}),
``How much time it takes for an algorithm to reach to \emph{its} near-optimal spread?'' (emphasis added by us). 
They define ``near-optimal'' in an empirical sense. We review their definition in Section~\ref{sec:ed}, where we also conduct an in-depth analysis.
Suffice it for now to note that the ``near-optimal'' spread is algorithm specific and can thus be drastically different for different algorithms. 
Thus, any comparison of the running times of different IM algorithms based on such an algorithm specific ``near-optimal'' spread necessarily holds the algorithms to different bars! To appreciate the seriousness of this issue, consider the following example. 

\begin{example}\label{ex:ed}
Consider two hypothetical IM algorithms $\A$ and $\B$. Suppose that 
$\A$ is an approximation algorithm that employs sophisticated sampling techniques, and thus the more samples it draws, the higher the quality of seed sets it can achieve.
Suppose the ``near-optimal quality'' of $\A$ corresponds to an estimated expected spread of $1000$ for $k=10$ seeds on some dataset $D$, and for that it takes 10 minutes. % to do so.
Algorithm $\B$ is a simple heuristic that takes 1 minute to get to {\sf its}  ``near-optimal quality'', namely a spread of $100$ for $k=10$ seeds on the same dataset $D$.
\eat{ If we require fewer samples for $\A$ so that it only achieves a spread of 10, $\A$ finishes in 0.1 minutes.} 
Suppose also that with fewer samples, Algorithm $\A$ achieves an estimated expected spread of $100$ and it takes only 0.1 minute to achieve this, again with 10 seeds. A comparison between Algorithms $\A$ and $\B$ solely based on the time taken by each to achieve its own ``near-optimal'' spread would conclude that $\B$ is a better algorithm despite the fact that $\A$ achieves what $\B$ achieves, an order of magnitude faster! 
\end{example}

In addition to the problematic issue of holding different algorithms to different bars, as demonstrated in Example~\ref{ex:ed}, there are two more problems associated with the experimental methodology of \cite{agr17}, as  discussed in Section~\ref{sec:ed}: $(i)$ As it turns out, the definition of ``near-optimal'' spread in~\cite{agr17} allows for a solution that can be {\sl arbitrarily worse} than the optimal spread, which calls into question the legitimacy of the term ``near-optimal'' spread; 
$(ii)$ No clear description is offered by Arora et al. \cite{agr17} on how exactly the parameters of the algorithms were set in order to obtain the ``near-optimal'' spread; as a result, it is unclear how each IM algorithm was allowed to run to reach such spread.

\spara{Measuring Standard Deviation} 
Apart from the flawed experimental design, other fundamental issues exist in their approach.
In Figure 12 of \cite{agr17}, Arora et al. reported a set of standard deviation values,
which are in turn used to determine the ``near-optimal quality'' of an algorithm.
%Needless to say, by the experimental design, 
By design, the correctness of these standard deviation measurements is critical to the correctness of their entire set of running time experiments.
However, in our attempts to reproduce their Figure 12, we found \emph{a significant discrepancy of 10x:
The standard deviation values obtained by us (validated at both NTU and UBC independently) are 10 times larger than the values claimed in Figure 12 of \cite{agr17}}.
Section~\ref{sec:ed} will explain how this issue has serious implications for  the validity of all other experiments in \cite{agr17}\footnote{In our email correspondences with them, the authors of \cite{agr17} stated that they performed ``binning'' and ``smoothing'' to remove outliers before calculating standard deviations. We reached out to them again to request the relevant source code for performing the exact same binning and smoothing so that we could reproduce their results. Despite multiple requests, we are still waiting for their source code. \eat{Unfortunately, even after multiple requests, they have not shared it with us.}}.

\spara{Algorithm-Specific Major Issues}
Arora et al. \cite{agr17} claimed to unearth several ``myths'' and criticized a variety of IM algorithms in the process of ``debunking'' those ``myths''. 
Unfortunately, many of such claims are false.
We shall refute them in detail in Section~\ref{sec:issues}. % and \ref{sec:misc}.
Here is an overview of those claims and our refutation. 

\begin{itemize}[leftmargin=*]
\vspace{-2mm}
\item 
%\eat{
%Stemming from the flawed design, Arora et al. hold the state-of-the-art approximation algorithm, IMM~\cite{tang15}, to an incredibly higher bar than any other algorithms, and through a series of logically erroneous arguments and experiments, Arora et al. claimed that IMM is no more efficient than TIM$^+$~\cite{tang14}, the very algorithm that IMM was proposed to improve (Section~\ref{sec:tim-imm}).
%} 
TIM and TIM$^+$ are the first scalable approximation algorithms for IM proposed by Tang et al. \cite{tang14} based on the notion of reverse-reachable sets~\cite{borgs14}.
In \cite{tang15}, Tang et al. proposed an improvement to TIM/TIM$^+$, called IMM, which leverages martingale theory to derive much tighter bounds on the number of samples required for a given guaranteed accuracy $\epsilon$ of spread estimation, compared to TIM/TIM$^+$.
IMM is much more scalable than TIM$^+$
	in the sense that for any given accuracy
	guarantee $\epsilon$, IMM requires far
	fewer samples to achieve that guarantee than TIM$^+$.

Arora et al. \cite{agr17} ignore the notion of theoretical accuracy guarantee and instead compare TIM$^+$ and IMM on \emph{empirical accuracy} and reach an erroneous conclusion that TIM$^+$ sometimes scales better than IMM. This reveals a lack of understanding of what theoretical guarantees are about, which are concerned with the worst case, as opposed to empirical accuracy.
In Section~\ref{sec:tim-imm}, we provide more arguments and details refuting their conclusion. To help appreciate the flaw in their scaling argument, we provide a simple example in Section~\ref{sec:tim-imm}. The example follows the same argument as used by \cite{agr17} and leads to the conclusion that Chebyshev's inequality is better than the Chernoff bound, which is clearly wrong!

\item Arora et al. claim that the SimPath algorithm~\cite{simpath} failed to finish on two datasets after 2400 hours (100 days!). A careful examination of their results and the datasets used reveals that they did not in fact preprocess the datasets correctly before running the SimPath code (released in \cite{simpath}), and as a result were stuck in an infinite loop. Without doing due diligence, they apparently simply let the code run for 100 days and concluded it does not finish in 100 days. 
Our experiments confirm that with correctly prepared data, SimPath takes only 8.6 and 667 minutes respectively to select 200 seeds on the two datasets tested, i.e., DBLP and YouTube (Section~\ref{sec:simpath}). 
One of the consequences of the incorrect results above is their claim that 
between the algorithms SimPath and LDAG, LDAG is more robust and is significantly faster under ``uniform'' LT model. As we discuss in Section \ref{sec:simpath}, our experiments contradict their claims. 

\vspace{-3mm}
\end{itemize}

\spara{Issues with Overall Recommendations by \cite{agr17}} 
In their paper, Arora et al. make a recommendation for which IM algoithm to use under what circumstances. This is expressed in the form of a decision tree (Figure 11(b) in \cite{agr17}). As a preview of the problems associated with this recommendation, notice that the decision tree recommends that whenever one has a low memory machine, the IM algorithm of choice is EasyIM \cite{easyim}.
%\pink{
We note that the EaSyIM algorithm was proposed by Galhotra, Arora, and Roy in SIGMOD 2016~\cite{easyim}. Both Galhotra and Arora are authors of \cite{agr17} as well. 
The rationale offered by them is that EasyIM is supposedly the most memory efficient IM algorithm and so should be used when available memory is limited. Indeed, as the authors point out, EasyIM stores only a scalar per node of the network, which requires only a fraction of the memory needed to store, e.g., reverse reachable sets. By this argument, the {\sl Random} algorithm, which randomly selects $k$ nodes from the network as seeds, requires even less memory since it does not need to store anything! However, it is well known that on many datasets, the expected spread achieved by the {\sl Random} algorithm is quite poor. What is missing in such a recommendation is a consideration of the trade-off between memory usage and spread achieved.
\eat{
\pink{Furthermore, according to \cite{agr17} itself, EaSyIM failed to finish on three largest datasets (out of eight) used in their benchmarking, and surprisingly, is recommended by Arora et al. as one of the four best IM algorithms.}
} 
We elaborate on the issues and misleading claims on EaSyIM~\cite{easyim} in Section~\ref{sec:misc}.

\subsection{Roadmap} 

The rest of our article is organized as follows.
%Section~\ref{sec:prelim} introduces more background knowledge on IM.
%Section~\ref{sec:algos} gives technical details on all algorithms that will be covered in the refutations. 
The background knowledge on IM as well as the algorithms covered in this article are covered in Appendix~\ref{sec:background}. 
Section~\ref{sec:ed} analyzes the %flawed 
experimental design and methodology of \cite{agr17} in detail and points out the flaws therein.
Section~\ref{sec:issues} focuses on refuting algorithm-specific mis-claims.
Section~\ref{sec:misc} discusses various other issues, including the problems associated with their claims on EaSyIM~\cite{easyim}. 
Appendix \ref{sec:celfpp} discusses the issue of CELF vs CELF++, while Appendix \ref{sec:app-unscientific} briefly remarks on the importance of understanding the source code of the algorithms that one undertakes to benchmark. 

 % and finally, Section~\ref{sec:concl} concludes this work.

%% file: sec-ed.tex
%\subsection{Evaluating Influence Maximization Algorithms} \label{sec:ed-eval}
\subsection{Standards of Benchmarking} \label{sec:ed-eval}

There exist several different metrics for evaluating an IM algorithm. Recall that some algorithms are just heuristics and lack guarantees on the expected spread of the seed set they return. We can only calibrate them on the empirical spread observed. For approximation algorithms, there is both empirical spread and the theoretical \emph{guarantee} on the worst-case spread. It is important to recognize the difference between these two. Besides spread, there are factors such as running time and memory usage. Any evaluation and comparison of IM algorithms should take into account the trade-off between resources like running time and memory on one hand and the achievable spread, i.e., empirical or guaranteed spread. Achieving higher spread usually requires larger computation costs in terms of running time and memory usage. 

\eat{ its theoretical worst-case guarantees, the quality of the seed sets it outputs (measured in terms
of the spread achieved),
%empirical accuracy\footnote{By
%accuracy of an algorithm, we mean the quality of seed sets produced by this
%particular algorithm. More precisely, it is the expected influence spread
%evaluated using sufficiently many MC simulations.}
the running time, and memory
consumption, etc. Among these metrics, there is a trade-off between (asymptotic
and empirical) spread achieved and the other others: achieving higher spread usually
requires larger computation costs and memory consumption, and so on.
}

Algorithms
often have one or more parameters to allow users to control this tradeoff, e.g.,
to reduce the computation time of the algorithm at the cost of spread, or vice
versa. For example, in case of Greedy with MC simulations, the \#MC simulations is a parameter that controls the trade-off between accuracy and runing time. In case of TIM$^+$ and IMM, the accuracy parameter $\epsilon$ can be used to trade accuracy for memory as well as running time. 
As a consequence, the setting of such parameters is crucial for benchmarking different algorithms carefully. 
% if one is to benchmark the performance of different algorithms.
%In this regard, a
A standard practice for parameter setting is to tune the parameters to \emph{ensure that all
algorithms are held to the same performance bar on one or more metrics, and then gauge
their performance on the other metrics}. 

\begin{figure}[t]
\centering
\includegraphics[width=.475\textwidth]{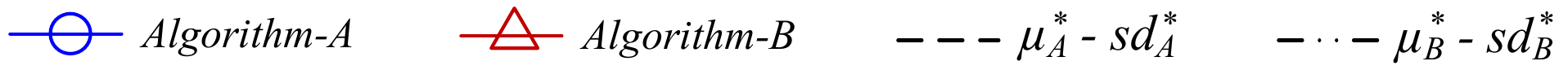} \\[-2mm]
%\begin{tabular}{cc} \hspace{-6mm}
%\includegraphics[width=.4\textwidth]{mock-plots/mock1.eps} & \hspace{-6mm}
%\includegraphics[width=.4\textwidth]{mock-plots/mock2.eps} \\ (a) The complete
%tradeoff curves.  & \hspace{-6mm} (b) Two data points.  \end{tabular}
\includegraphics[width=.33\textwidth]{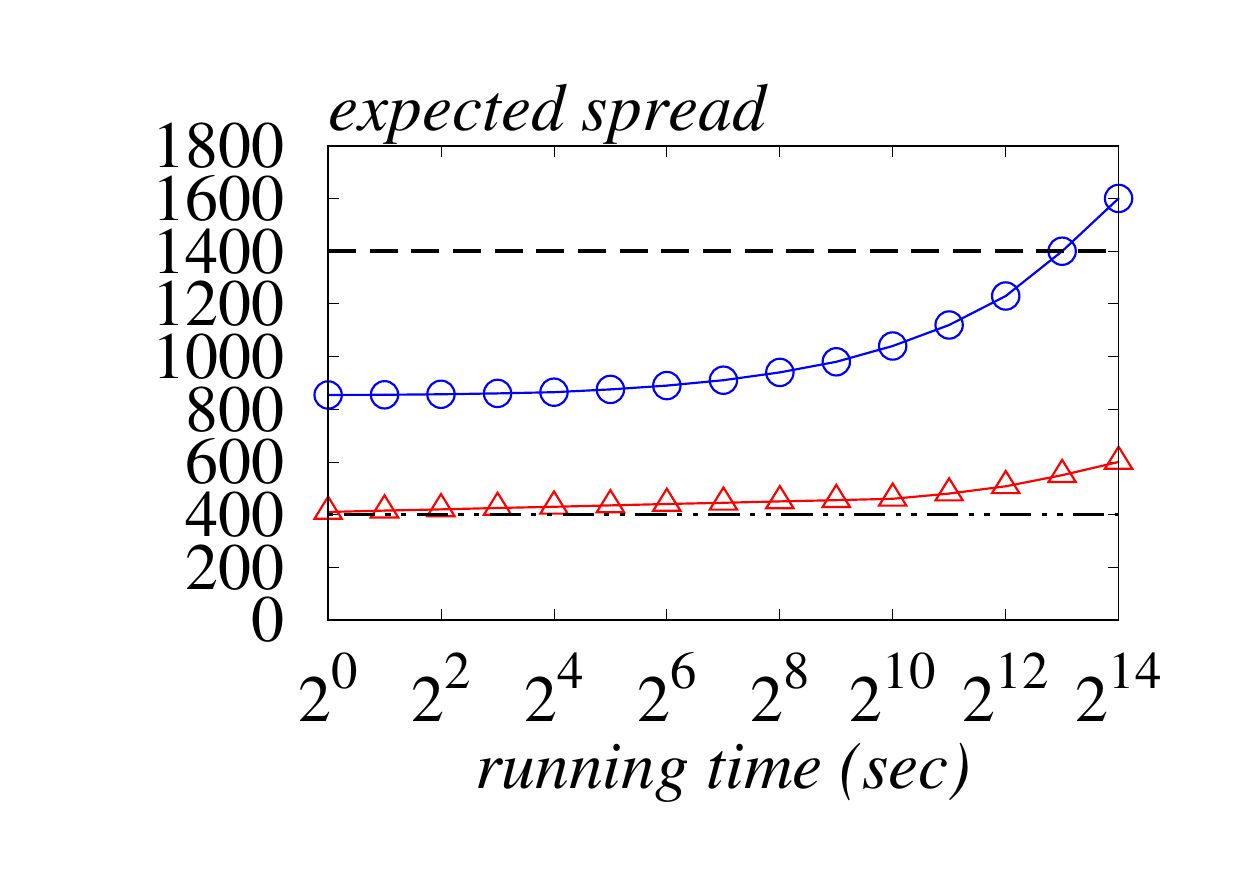} \vspace{-2mm}
\caption{The tradeoff between expected spread and running time for two
conceptual algorithms on a certain dataset, with seed set size $k = 200$.}
\label{fig:ed-mock1}
\end{figure}

Suppose that for some reason, we wish to deviate from the standard practice and
choose an alternative methodology for benchmarking IM
algorithms, then we must ensure that the alternative
methodology satisfies some minimum requirements for soundness. 
Consider Figure~\ref{fig:ed-mock1} for an example.
It illustrates the expected spreads of two
hypothetical IM algorithms when their running time varies (due to parameter tuning),
for a given dataset and a seed set size $k=200$. Observe that, in terms of the
tradeoff between spread achieved and computation cost, Algorithm $\A$ completely
dominates Algorithm $\B$, since the former $(i)$ yields a higher expected spread when
the two algorithms have the same computation time, and (ii) incurs a smaller
running time when the two algorithms offer the same expected spread. As
such, \emph{if a methodology for evaluating IM algorithms is
sound, then it should not conclude that Algorithm $\B$ is more
favorable than Algorithm $\A$ in terms of running time or spread achieved when $k = 200$}.
Unfortunately, this is the very conclusion that the methodology of \cite{agr17} would draw, and hence
%their experimental methodology 
it does not pass this soundness check, as we elaborate below.

\subsection{Flaws in Arora et al.'s Methodology} \label{sec:ed-flaw}

%\spara{Methodology of \cite{agr17}}
Essentially, the experimental methodology of \cite{agr17} is driven by the following question: {\em How much time it takes
for an algorithm to reach to {\bf its} near optimal spread?} To answer this question,
%In a nutshell,
Arora et al.\ choose a {\em fixed} parameter setting for each influence maximization algorithm
$A$, and then evaluate $A$ on three metrics: spread achieved, computation time,
and memory consumption.  Specifically, if algorithm $A$ has a parameter $p$,
then they first run $A$ on four small datasets, with a fixed seed set
size $k$ ($k$=200 in their settings), and with $p$ varying in a certain range.
Let $\mathcal{S}_D$ be the set of seed sets returned by $A$ on a dataset $D$ for
different $p$. Arora et al.\ examine the expected spread of each seed set in
$\mathcal{S}_D$ using $10000$ rounds of Monte-Carlo simulations, and identify
the seed set $S^*_D$ whose expected spread $\mu^*_D$ is the largest.
In addition, they measure the standard deviation $sd^*_D$ of the spread of
$S^*_D$ in the above $10000$ Monte-Carlo rounds. The final parameter value $p$
is set to a value that minimizes the computation of $A$, subject to
the constraint that $A$ should achieve an expected spread at least $\mu^*_D -
sd^*_D$, across all four small datasets. In other words, they set $p$ in a
manner such that $A$'s spread achieved is close to \emph{its} ``best'' (or, ``near-optimal'').
It is worth noting that once $p$ is chosen for an algorithm, {\sl it remains the same across all datasets, including the bigger ones.}

\subsubsection{Issue 1: Different and arbitrary bars for different algorithms}
\label{sec:ed-issue1}
Simply put, the ``best'' expected spread is algorithm specific and can be considerably different. \emph{By construction, then different algorithms are not held to the same bar w.r.t. expected spread.}  
%are {\sl not equally good}. 
If an algorithm's best expected spread is higher than
another, then Arora et al.'s methodology would require the former to achieve a
higher spread than the latter when it evaluates the two algorithms' running
time and memory consumption. Such a setting unfairly penalizes algorithms whose
best expected spreads are large, and could lead to erroneous conclusions
regarding their efficiency and memory overheads, as well as wrong recommendations for algorithms to use for given situations. 
In other words, the fact that an algorithm is capable of achieving a higher
	quality of seed sets will be used against it in running time comparisons!

%The problem here is that

%so-called ⤽best⤝ seed sets of different methods are not equally good (in terms
%of the influence spread metric). As just one example, in Figure 13 of [1], the
%⤽best⤝ 200-seed set for IMM under the IC model has an expected spread of almost
%4250, whereas the ⤽best⤝ seed set for IMRank [3] under the same setting has an
%expected spread around 4000. In that case, it does not make sense to compare
%the running time of IMM and IMRank since the quality of their outputs is
%different. In fact, if we only require IMM to achieve an expected spread of
%4000, then it would be two orders of magnitude faster than the case when it
%yields an expected spread of 4250.

In the context of Figure~\ref{fig:ed-mock1}, where $k = 200$, Algorithm $\A$
(resp.\ Algorithm $\B$) achieves its largest expected spread $\mu^*_A = 1600$
(resp.\ $\mu^*_B = 600$) when its running time equals $2^{14}$ seconds. Assuming
that $sd^*_A = sd^*_B = 200$, then Arora et al.'s setting would require
Algorithm $\A$ to yield an expected spread of at least $\mu^*_A - sd^*_A = 1400$,
in which case its running time is at least $2^{13}$ seconds. In contrast,
Algorithm $\B$ is only required to achieve an expected spread of at least $\mu^*_B
- sd^*_B = 400$, in which case its computation time can be as small as $1$
second. Evaluating the two algorithms' efficiency under such a setting is
clearly unfair to Algorithm $\A$, and could lead to {\em incorrect} conclusions such
as the following:
%In other words, for $k = 200$, Arora et al.'s experimental setting would ignore
%the accuracy-efficiency tradeoff curves shown in
%Figure~\ref{fig:ed-mock1}, and would only focus on the two data points
%shown in Figure~\ref{fig:ed-mock1}b. This could lead to completely
%incorrect conclusions, such as:
\begin{enumerate}[leftmargin=*]
\vspace{-1mm}
\item Algorithm $\A$ is not scalable due to its excessive
computation time needed to achieve a spread of 1400, even though this is more than 3x the spread 400 expected of Algorithm $\B$. 
\vspace{-2mm}
\item Algorithm $\B$ is more efficient than Algorithm $\A$, despite the fact that Algorithm $\A$ can achieve the same spread of 400 in lower running time! 
%that its accuracy
%is lower than the latter's.
\vspace{-1mm}
\end{enumerate}

%Observe that 
Both conclusions clearly contradict the fact that Algorithm $\A$ dominates Algorithm $\B$,
as depicted in Figure~\ref{fig:ed-mock1}.

Now take a concrete example: In Section 5.3.1 of \cite{agr17},
Arora et al.\ conclude that IMM is not scalable under the IC model.
However, this conclusion is drawn solely based on IMM's running time when it is
required to achieve an extremely high accuracy (with its parameter $\epsilon$
set to $0.05$), as Arora et al.\ experimental methodology demands that IMM
should yield an expected spread close to \emph{its} best. Meanwhile, other methods
(e.g., EaSyIM) are not held to the same bar of spread, since their best
expected spreads are lower than that of IMM. If we allow IMM to output less
accurate results as other methods do, then it would become much more efficient and scalable.

\begin{figure}[t!]
\centering
\vspace{-3mm}
\includegraphics[width=.33\textwidth]{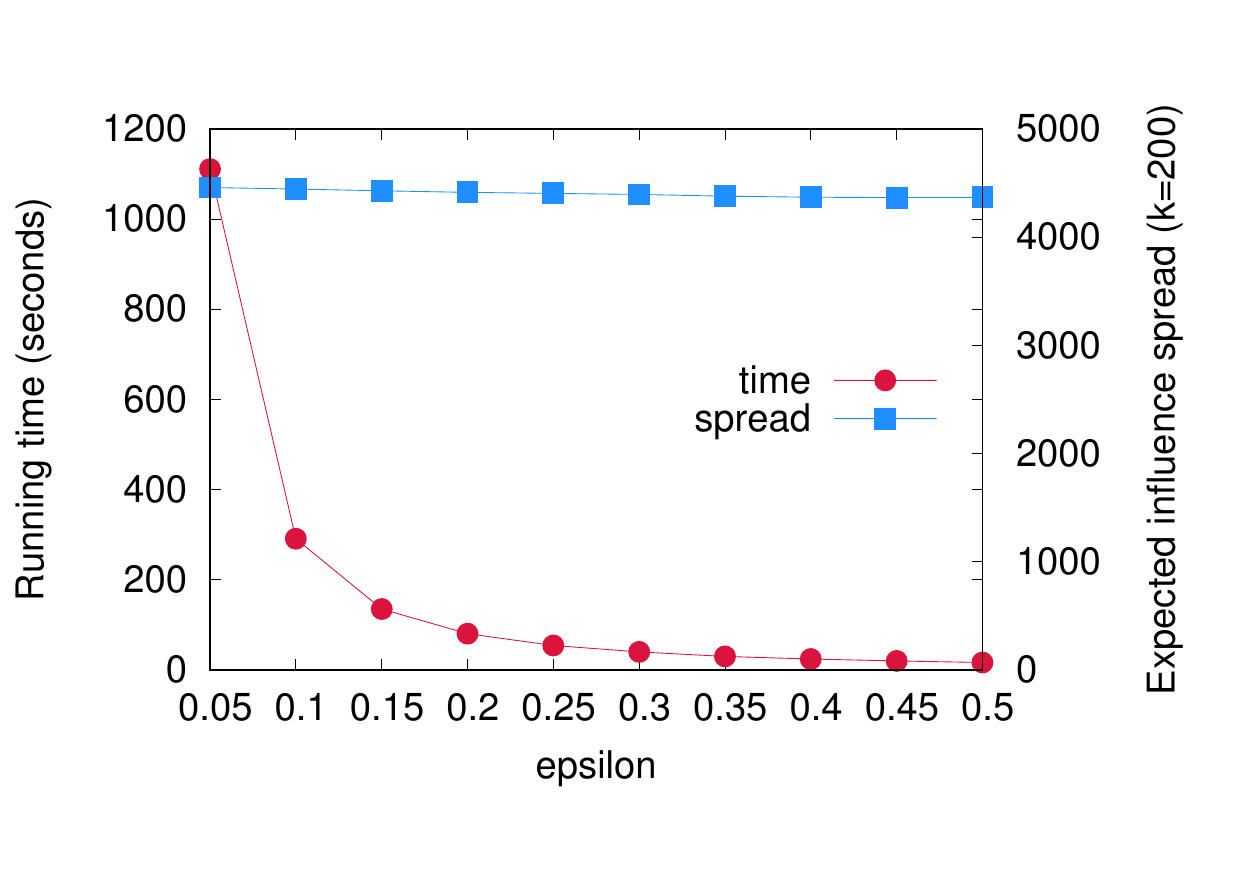}
\vspace{-6mm}
\caption{IMM on HepPH-IC: Running time (left $Y$-axis) and spread at $k=200$ (right $Y$-axis), with varying $\epsilon$.}
%\vspace{-3mm}
\label{fig:imm-tradeoff}
\end{figure}

For instance, consider Figure~\ref{fig:imm-tradeoff}.
It depicts  results of our experiments on IMM's running time and spread achieved (at $k=200$) on HepPH dataset, under the IC model, with $\epsilon$ varied over $0.05, 0.1, \ldots, 0.5$.
Clearly, as $\epsilon$ increases, {\sl the drop of spread is barely visible, while the running time goes down drastically.} 
We can see that $\epsilon = 0.05$, the choice in \cite{agr17}, is almost an extreme adversarial setting for IMM: It took significantly more time, but produced only marginally better spread than larger $\epsilon$'s.
For instance, \emph{from $\epsilon=0.05$ to 0.5, the running time speed-up is about 68x, while the drop in spread is only 2.1\%!} Similar observations hold for other datasets too. It follows that without sacrificing spread too much, it is possible to set $\epsilon$ to higher values, thus \emph{easily} achieving scalability.

Hence, Arora et al.'s claim that  IMM does not scale is invalid,
and it is a consequence of their flawed experimental methodology.

\begin{figure}[t]
\centering
\includegraphics[width=.35\textwidth]{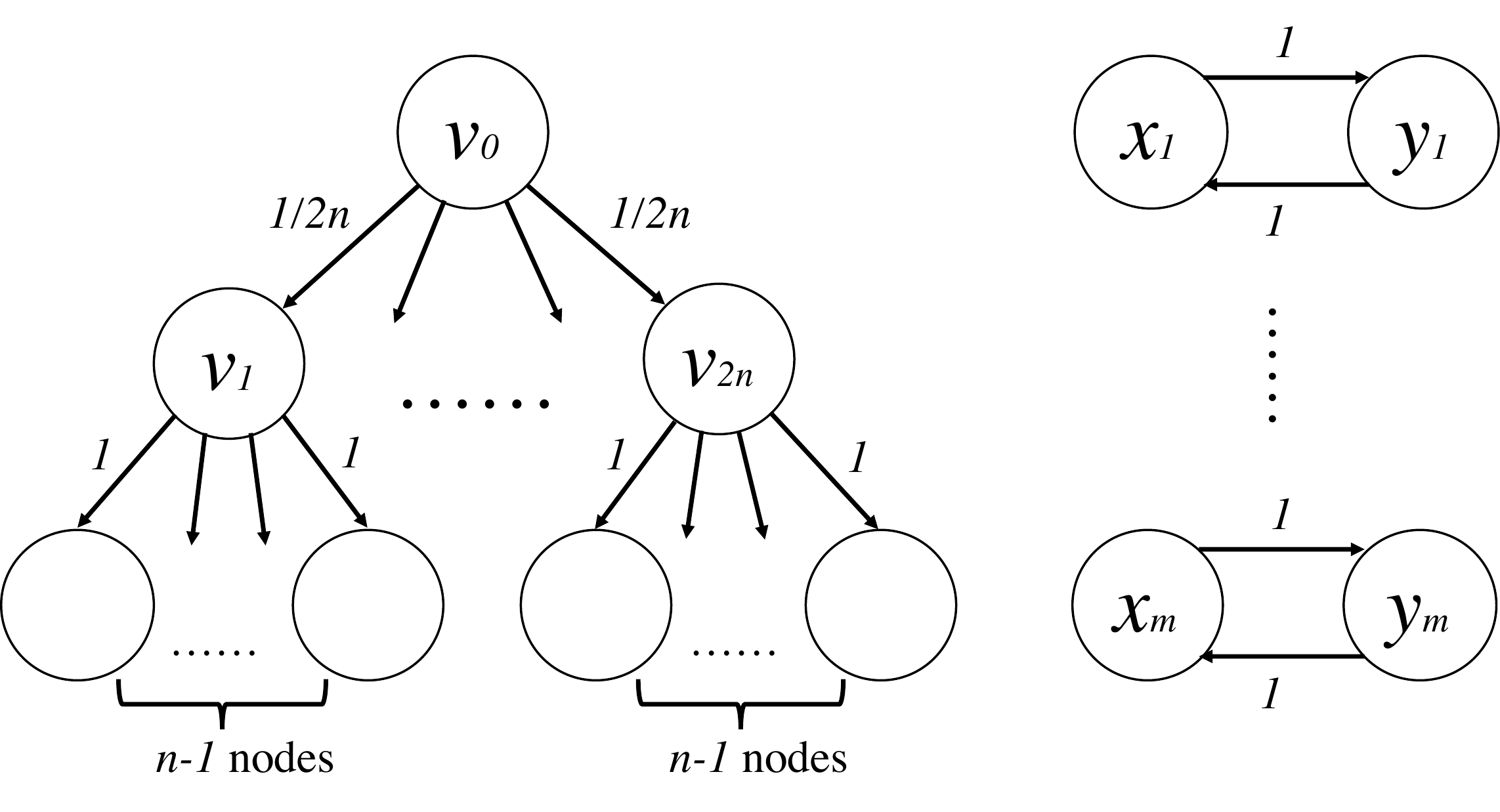} \vspace{-2mm}
\caption{A counter example showing $\mu^*-sd^*$ giving arbitrarily bad spread. In the example graph, we let $m = n^3$. (Credit: Yang and Pei~\cite{yu-pei17})}
\label{fig:mu-std-ex}
\end{figure}

\subsubsection{Issue 2: How close to optimal is ``near-optimal'' spread?}
Arora et al.~\cite{agr17} define the ``near-optimal'' influence spread using $\mu^*-sd^*$, as mentioned in the beginning of Section~\ref{sec:ed}.
This, unfortunately, is an ad-hoc choice and we show through a counterexample contributed by Yu Yang and Jian Pei~\cite{yu-pei17} that the value of $\mu^*-sd^*$ can be {\em arbitrarily} worse than the optimal influence spread.
Consider Figure~\ref{fig:mu-std-ex}, where the graph is constructed as follows.
Let $n \ge 2$ be a positive integer.
There is a node $v_0$ that has $2n$ outgoing edges, pointing to $v_1, v_2, \ldots, v_{2n}$ each with an influence probability of  $\frac{1}{2n}$.
Each $v_i$ ($i \in [1,2n]$) has $n-1$ out-neighbors each with influence probability $1$.
Then we create $m = n^3$ 2-cliques: $\{x_1, y_1\}, \{x_2, y_2\}, \ldots, \{x_m, y_m\}$.
 Each pair $\{x_j, y_j\}$ forms a 2-clique with influence probabilities being $1$ on both edges.

Let $I(S)$ denote the number of active nodes by the end of a propagation when $S$ is the seed set.
Clearly, by definition, $\mathbb{E}[I(S)] = \sigma(S)$.
For the example graph in Figure~\ref{fig:mu-std-ex}, under the independent cascade model, when $k=1$, it can be verified that the following holds:
\begin{subequations}
\begin{align}
\mu &:= \mathbb{E}[I(\{v_0\})] = \sigma(\{v_0\}) = 1+n , \label{eqn:mu-opt} \\
sd &:= \sqrt{\mathit{Var}[I(\{v_0\})]} = n \sqrt{1-\frac{1}{2n}} ,   \\
\mu - sd &= 1+n \left(1- \sqrt{1-\frac{1}{2n}} \right) \leq 1 + n \frac{1}{2n} = 1.5. \label{eqn:mu-std-constant}
\end{align}
\end{subequations}

On the other hand, for all $2n^3$ nodes in the 2-cliques structure, their expected spread is $2 > 1.5 = \mu - sd$.
If we randomly select a node from this graph, then w.h.p.\ we will end up with a node of spread $2$.
Arora et al.~\cite{agr17} use $\mu^* - sd^*$ as the threshold for ``near-optimal'' spread.
Since $\mu^*$ and $sd^*$ are empirical estimates of $\mu$ and $sd$, one of the following scenarios must hold true if their approach is followed:
\begin{enumerate}[leftmargin=*]
\item $\mu^*$ and $sd^*$ are reasonably close to $\mu$ and $sd$ respectively. %to the latter quantities.
Thus, $\mu^* - sd^*$ should be reasonably close to $\mu-sd$ which is bounded by a small constant 1.5 (Eq~\eqref{eqn:mu-std-constant}).
As a result, the threshold $\mu^* - sd^*$ used for defining ``near-optimal'' spread can be arbitrarily smaller than the true optimal spread $\mu$ since it has a $\Omega(n)$ gap to the latter. This invalidates the use of $\mu^*-sd^*$ as the threshold for defining ``near-optimal'' spread.

\item The quantity $\mu^* - sd^*$ is reasonably close to the optimal spread $\mu$, thus justifying a definition of ``near-optimal'' based on the threshold $\mu^* - sd^*$. However, in this case, one or more of the quantities in $\{\mu^*, sd^*\}$ must be a poor estimate of their counterparts in $\{\mu, sd\}$. This, by definition, calls into question the accuracy of $\mu^*$ and/or  $sd^*$ itself, and thus the validity of all experimental results based on such poor estimates will be questionable at best.
\end{enumerate}

\eat{
which would be regarded ``near-optimal'' according to Arora et al.~\cite{agr17}.
However, since the true optimal spread is $n+1$ (Eq.~\eqref{eqn:mu-opt}), and hence the so-called ``near-optimal'' spread has a {\em linear} $\Omega(n)$ gap to the true optimal.
This establishes that the ``near-optimal'' spread in Arora et al.~\cite{agr17} is mathematically ill-defined, and thus any results or conclusions drawn from their experiments are highly unlikely to be scientifically worthy.
}

\subsubsection{Issue 3: Notion of reasonable time limit is arbitrary and unclear}
In addition to the fundamental issues mentioned above, Arora et al.'s methodology suffers from yet another critical issue.
Recall that while choosing the value of parameter $p$ (which is used to control the trade-off between running time and expected spread), Arora et al.\ allow algorithms to run for a ``reasonable time limit'' (Section 5.1.1 of \cite{agr17}). 
However, unfortunately, no  definition or value of reasonable time limit for any of the studied algorithms is  provided. This makes the notion arbitrary and unclear, and worse, leaves ``reasonable time limit'' open to interpretation. Clearly, as more time is allowed, the parameter $p$ gets stricter: e.g., more MC simulations or smaller accuracy parameter $\epsilon$. In a fair world, a strict upper bound $T$ on running time would have been provided, for {\em all} algorithms, making reasonable time limit clear and concrete. 
In other words, not only Arora et al.'s experimental methodology sets the bars unfairly for different algorithms (as described in Section \ref{sec:ed-issue1}), even the method for computing the bars (aka, ``near-optimal'' quality of each algorithm) is ill-defined.

%Clearly, if an algorithm $A$ is allowed to run for time $T_A$, while algorithm $B$ is allowed to run for time $T_B$, where $T_A > T_B$, then the parameter
%\pink{
To appreciate the gravity of this issue, consider a thought experiment where we have two exact replicas $\A_1$ and $\A_2$ of the same algorithm $\A$. Assume that Arora et al.'s experimental methodology is not aware that $\A_1$ and $\A_2$ are essentially the same algorithm. 
While tuning their parameters $p$ for the two algorithms, if the same upper bound on running time $T$ is not employed, then we will end up with two different values of $p$, namely $p_{\A_1}$ and $p_{\A_2}$ for the two replicas, respectively. W.l.o.g., suppose $p_{\A_1}$ is stricter than $p_{\A_2}$ (e.g., smaller accuracy threshold $\epsilon$). As a result, $\A_1$ and $\A_2$ are necessarily held to different bars on spread and hence we may conclude that one of them is more efficient than the other. However, since $\A_1$ and $\A_2$ are replicas of each other, such a conclusion is absurd. 

\eat{ 
W.l.o.g., assume $p_B$ is stricter than $p_A$ (e.g., smaller $\epsion$). That is, algorithm $\A_1$ is allowed more time than $\A_2$ during the selection of their parameters. Therefore, $\A_1$'s near optimal spread is now considered higher than $\A_2$'s near optimal spread and as a result, $\A_1$ is now held against a higher bar than $\A_2$, on all datasets\footnote{Recall that once the parameter value is picked, it is kept the same across all the datasets.}.
Eventually, Arora et al.\ will conclude that $\A_2$ is more efficient and scalable than $\A_1$, even though they are exact replicas of the same algorithm. 
} 
%if $A_1$ is given more time than $A_2$, then the value of $p$ for algorithm $A_1$ could be smaller than $A_2$ (lower value of $p$ implies higher accuracy), and as a result, $A_1$ is now held against a higher bar than $A_2$, on all datasets\footnote{Recall that once the parameter value is picked, it is kept the same across all the datasets.}. Even a small difference in the values could have an amplified differences in running times on much bigger datasets. Eventually, Arora et al.\ will conclude that $A_2$ is more efficient than $A_1$, even though they are exact replicas of the same algorithm.
%}

\eat{ 
\note[Wei]{@laks, Amit's thought is that they did not say anything concrete about the time limit. Hence, the answer to ``how to decide an algorithm's near optimal spread'' remains unclear from their paper, and readers can draw arbitrary interpretation and conclusions.  This pink example is one example.  Please see if you can refine his points better?}
} 

\subsection{Irreproducible Results in Arora et al.'s Parameter Setting}
\label{sec:ed-exp}
Furthermore, we are unable to reproduce a vital experiment in
\cite{agr17} that was used to determine the parameter of each algorithm.
We now elaborate on this.

\begin{figure*} \centering
\begin{tabular}{ccc} \hspace{-2mm}
\includegraphics[width=.3\textwidth]{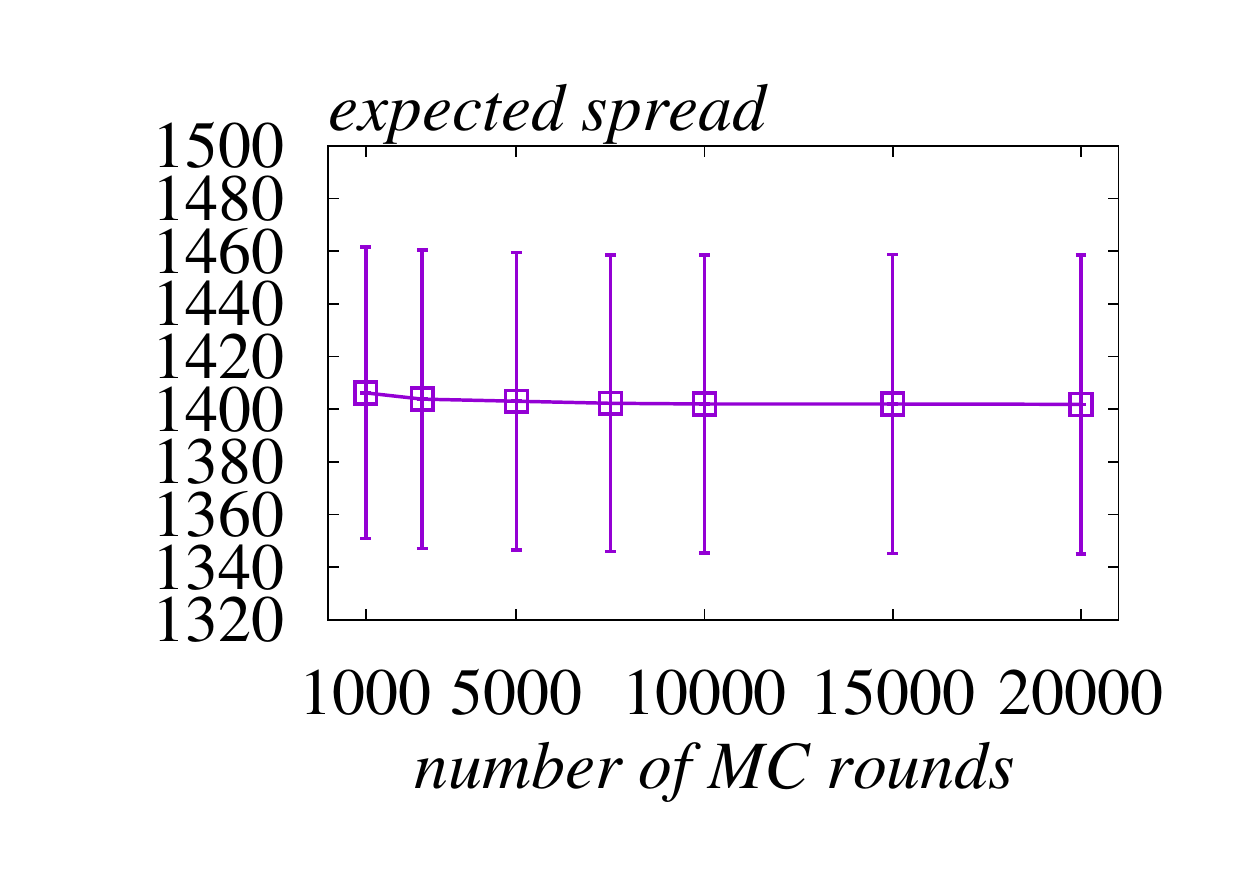} & \hspace{-5mm}
\includegraphics[width=.3\textwidth]{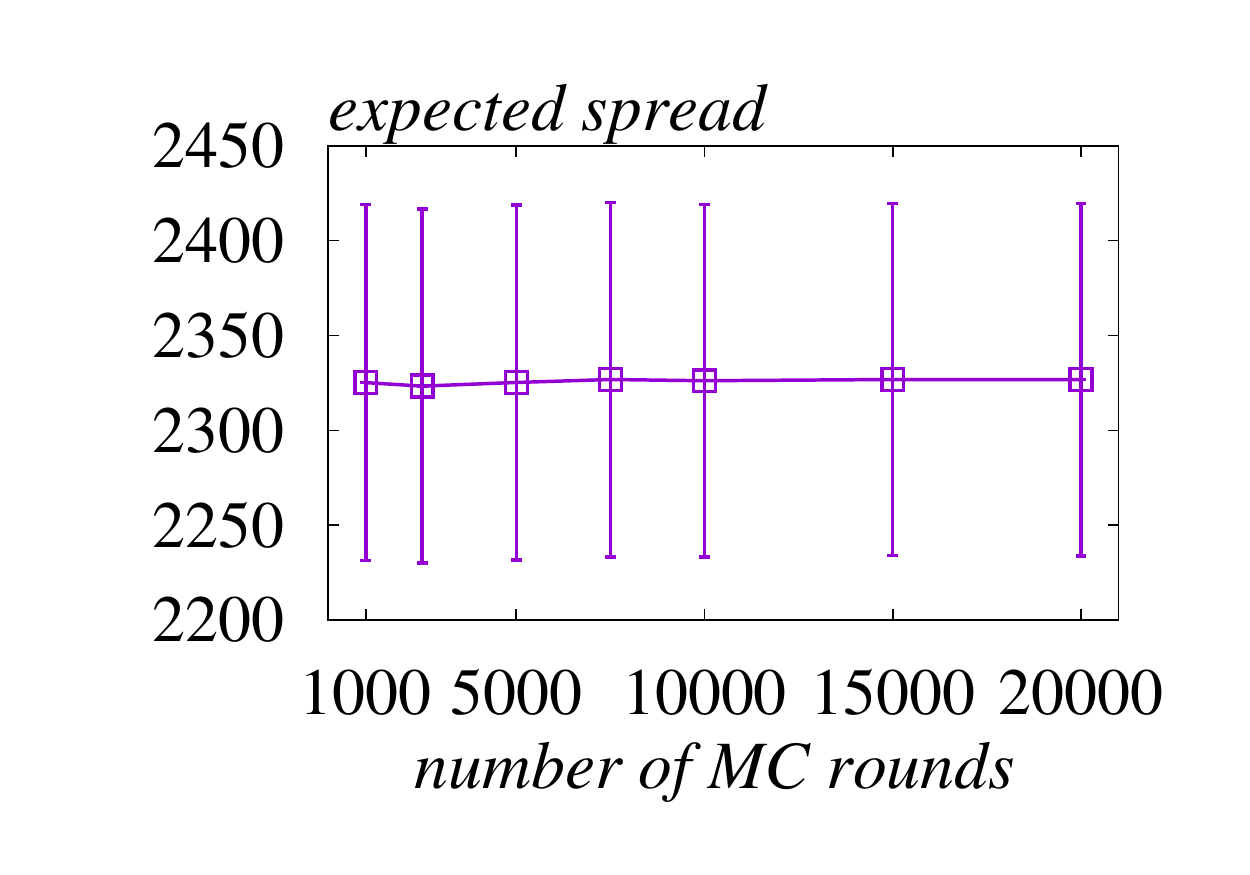} & \hspace{-5mm}
\includegraphics[width=.3\textwidth]{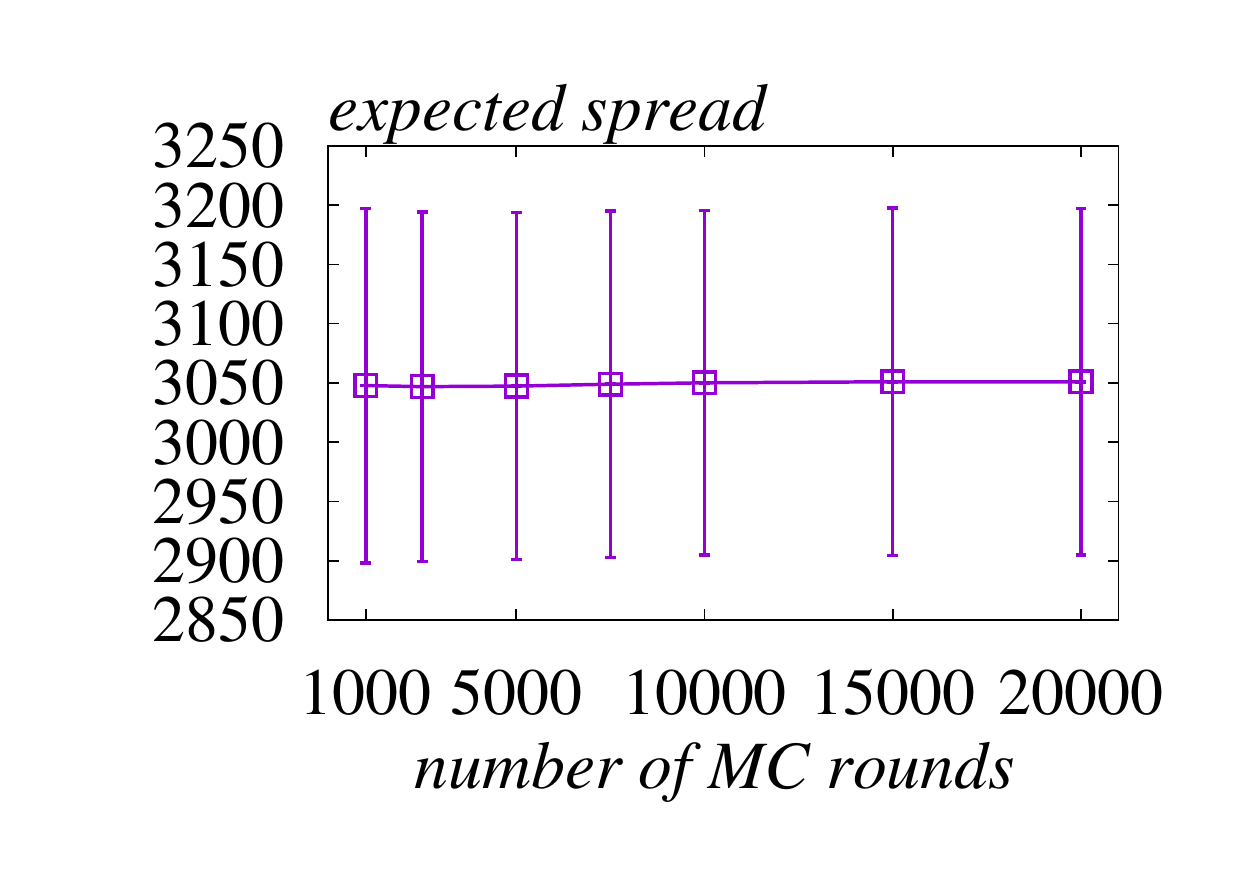} \\[-2mm] (a)
Nethept (IC).  & \hspace{-8mm} (b) Nethept (WC). & \hspace{-8mm}  (c) Nethept
(LT).\\%[2mm] \hspace{-2mm}
\includegraphics[width=.3\textwidth]{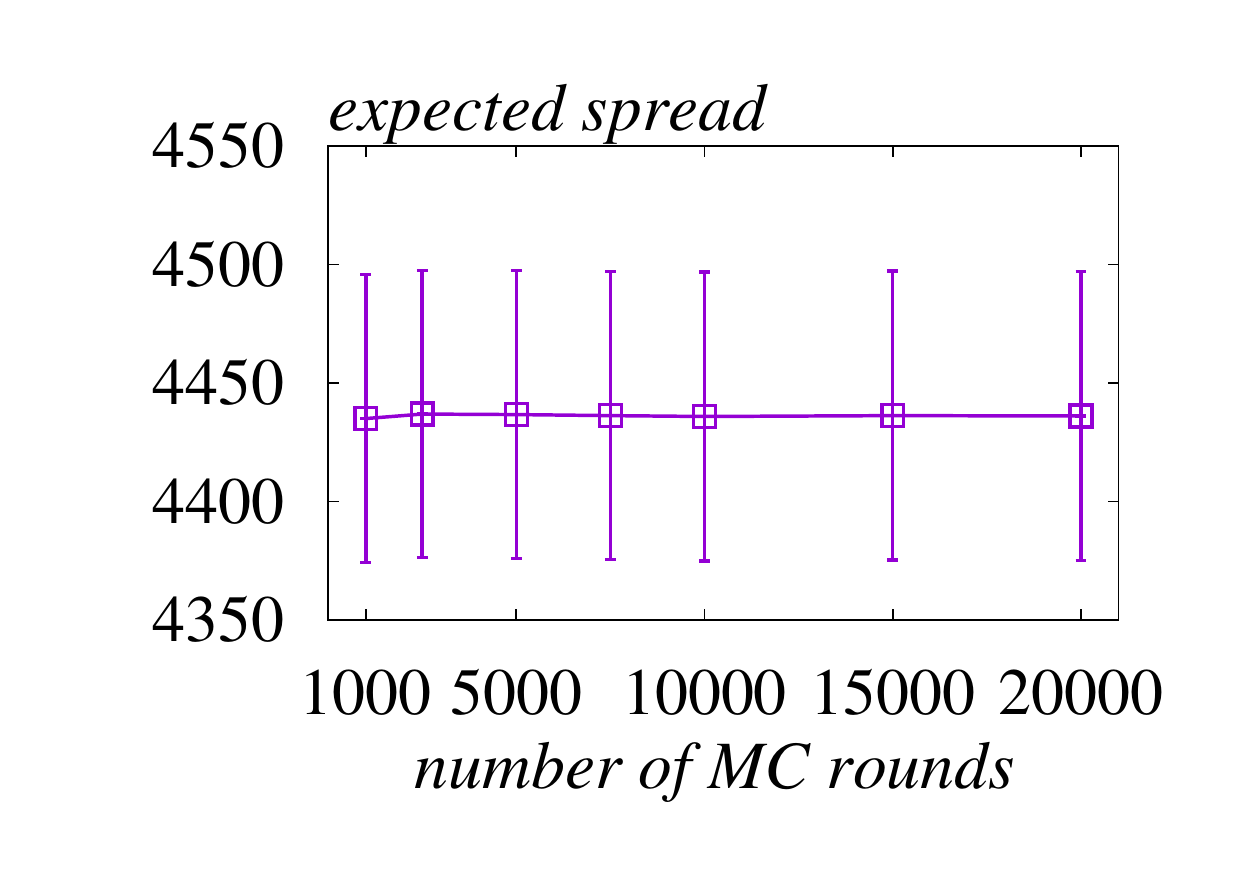} & \hspace{-5mm}
\includegraphics[width=.3\textwidth]{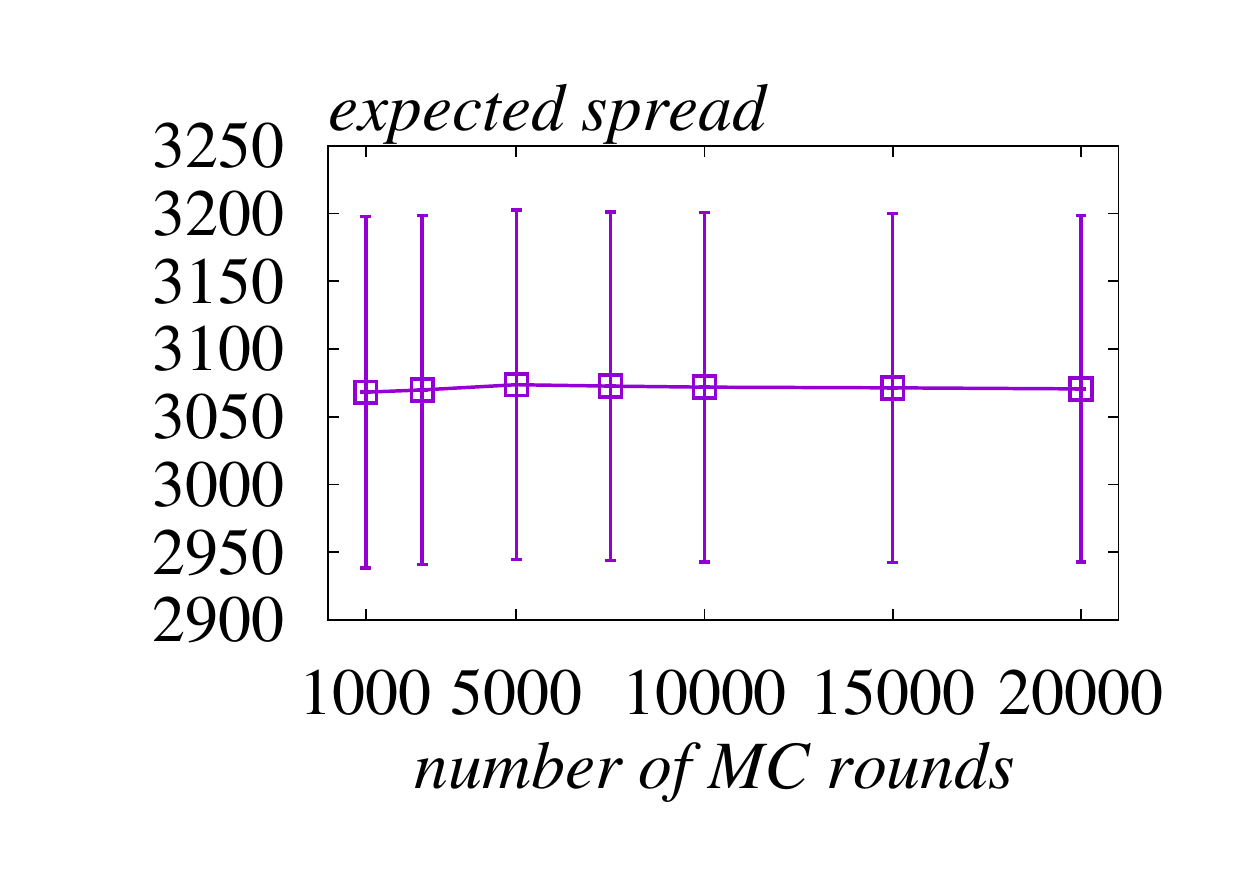} & \hspace{-5mm}
\includegraphics[width=.3\textwidth]{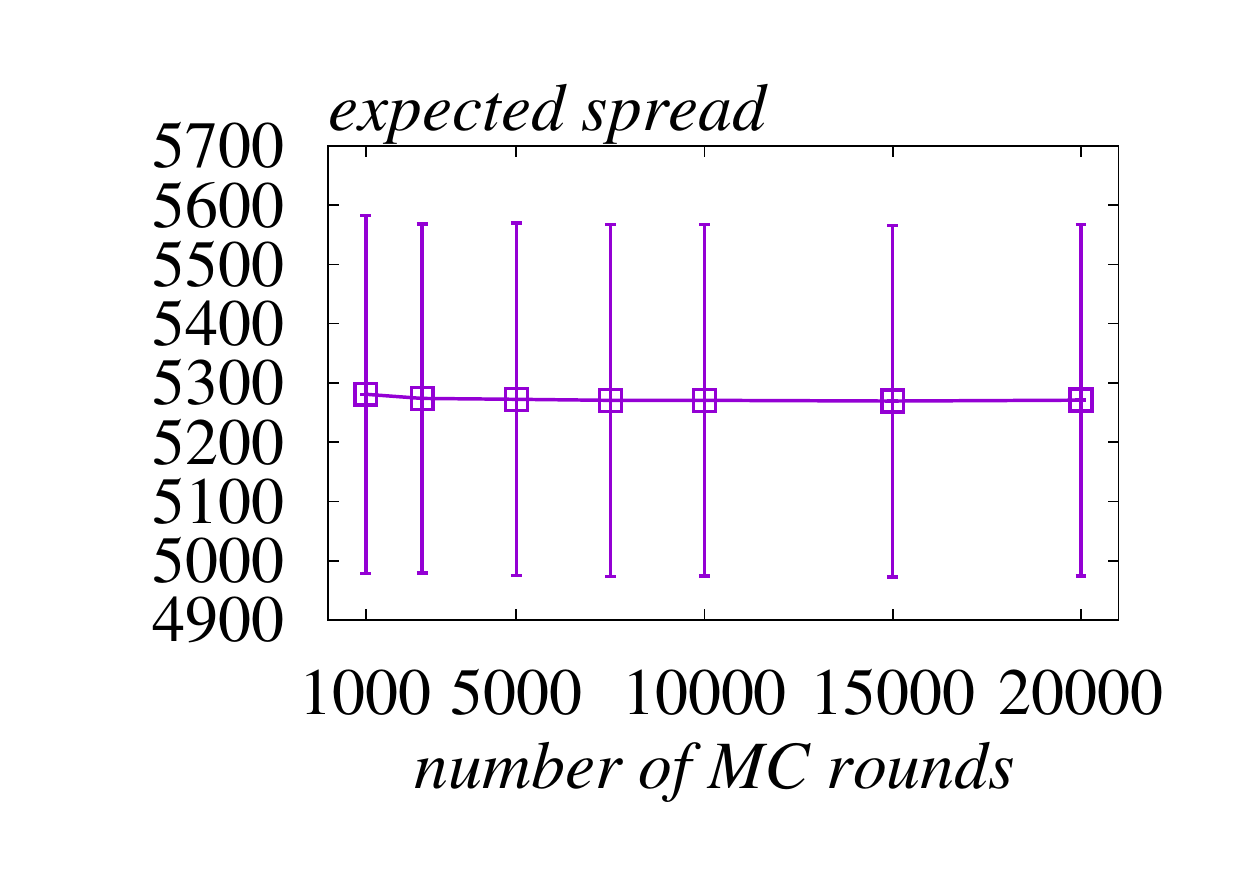} \\[-2mm] (d) HepPH
(IC).  & \hspace{-8mm} (e) HepPH (WC). & \hspace{-8mm}  (f) HepPH (LT).
\end{tabular} \caption{$\mu^*$ and $sd^*$ on Nethept and HepPH.}
\label{fig:ed-std} \end{figure*}

Recall that Arora et al.\ set the parameter $p$ of each algorithm $A$ such that
it yields a seed set whose expected spread is at least $\mu^* - sd^*$, where
$\mu^*$ is the expected spread of the best seed set $S^*$ that $A$ can produce
over a range of values of $p$, and $sd^*$ is the standard deviation of $S^*$'s spread in
$10000$ rounds of Monte-Carlo simulations. In other words, Arora et al.'s
parameter setting for $A$ highly depends on the measurement of $\mu^*$ and
$sd^*$. In Figure 12 of \cite{agr17}, Arora et al.\ report their measurements of
$\mu^*$ and $sd^*$ for IMM, using seed set size $k=200$ and varying the number
of Monte-Carlo rounds from $1000$ to $20000$. Values of $sd^*$ for other algorithms are not reported in their paper\footnote{We requested the authors to share their code that computes standard deviations. Unfortunately, even after several attempts, we are still waiting for their code. 
\eat{the authors have yet to respond to our requests.} }.

We attempted to reproduce Arora et al.'s results on two of the datasets that they
use, i.e., Nethept and HepPH. On each dataset, we used their setting and
ran IMM with $\epsilon = 0.05$ and $k = 200$ to obtain $S^*$, and then measured  $\mu^*$ and $sd^*$ using $r$
rounds of Monte-Carlo simulations, with $r$ varying from $1000$ to $20000$.
%In particular,
We compute $\mu^* = \frac{1}{r}\sum_{i=1}^{r} I_i\left(S^*\right),$
where $I_i\left(S^*\right)$ denotes the spread of $S^*$ in the $i$-th
Monte-Carlo round.
In addition, we compute $sd^*$ using the standard formula for
sample standard deviation:
\begin{align} \label{eqn:ed-std}
sd^* =
\sqrt{\frac{1}{r-1}\sum_{i=1}^{r} \left(I_i\left(S^*\right) - \mu^*\right)^2}  \,
\end{align}

Figure~\ref{fig:ed-std} illustrates the values of
$\mu^*$ and $sd^*$ that we obtain from each dataset under the IC, WC, and LT
models. On Nethept, the values of $sd^*$ under the IC, WC, and LT models are
approximately 50, 90, and 150, respectively, while on HepPH, the values of $sd^*$
under the IC, WC, and LT models are roughly 60, 120, and 300, respectively. In
contrast, the values of $sd^*$ reported in Figure 12 of \cite{agr17} (for 10000
Monte-Carlo rounds) are around {\sl 10 times} smaller than the corresponding values in
Figure~\ref{fig:ed-std}. For instance, in case of the IC model on Nethept, Arora et
al.\ report that $sd^* < 5$. Such a small $sd^*$ is anomalous given the large
amount of randomness in Monte-Carlo simulations.

We note that such significant errors (10x gap)
in the measurement of $sd^*$ {\sl have serious implications on
all experimental results in \cite{agr17}}: If $sd^*$ is
measured incorrectly, then the parameter settings for all experiments in
\cite{agr17} are erroneous, since the latter are decided based on the former. To
correct such serious errors, all experiments in \cite{agr17} would have to be
re-run. We elaborate further on this in the next para. 

\spara{Reasons for Discrepancy}
We corresponded with Arora et al.\ via email regarding the above issue on $sd^*$, and were told that the discrepancy is due to the fact that they performed ``binning'' and ``smoothing'' operations on the distribution before calculating $sd^*$.
They claimed that such operations were done to ``remove outliers''. It is noteworthy that there is no mention of binning and smoothing in the paper \cite{agr17}. We requested them for a detailed explanation of how they performed binning and smoothing and/or the code they used for performing these operations. Despite multiple reminders, we \eat{have yet to hear from them.} are still waiting for their code and explanations.

Regardless of how this ``binning'' was conducted, we note that there are at least three serious issues at hand here.
First, Eq~\eqref{eqn:ed-std} is the standard definition of sample standard deviations, and thus conducting any ``binning'' operations prior to the application of  Eq~\eqref{eqn:ed-std} and still calling it ``standard deviation'' (i.e., without any qualification) is misleading at best. Second, the notion of ``outliers'' is irrelevant when measuring standard deviations because, for any random variable $v$ sampled from a distribution $\Omega$, the standard deviation of $v$ is defined over all possible samples from $\Omega$ -- that is, by definition, no sample from $\Omega$ should be ruled as outliers. Third, in the paper itself \cite{agr17}, {\sl there is no mention of the ``binning'' operations at all}, which is {\sl rather strange} given that they have such profound effects in the measurement of $sd^*$. 

For instance, if no ``binning'' or ``smoothing'' is done and
$sd^*$ of IMM is taken from our Figure~\ref{fig:ed-std}(d), then 
IMM would be allowed to run with a spread bar of 4380.
From Figure~\ref{fig:imm-tradeoff}, this implies that $\epsilon$ can be set at $0.35$,
with running time 29.8 sec, a 37.2x speed-up compared to $\epsilon=0.05$!

%% file: sec-tim-imm.tex
\begin{misclaim}
\textit{``Both TIM$^+$ and IMM do not scale beyond HepPh in terms of memory-consumption under the IC model.''}
-- \cite{agr17}, Section 5.2, Section 5.3.1, Section 5.4 (repeated).
\end{misclaim}

\noindent\textbf{Refutation:}
Arora et al.\ make this claim solely based on the running time of TIM$^+$ and IMM when they set $\epsilon = 0.05$. This setting, as we mentioned in Section~\ref{sec:ed-exp}, requires TIM$^+$ and IMM to generate extremely accurate results, which significantly increases the computation overheads of both algorithms. In contrast, other algorithms compared (e.g., EaSyIM) are allowed to produce less accurate results (due to the experimental methodology used in \cite{agr17} -- see Section~\ref{sec:ed-eval}), which leads to much smaller computation costs. If we lower the accuracy of IMM to the same level as other algorithms (e.g., EaSyIM), then it would become much more scalable. Therefore, the above claim is invalid.
%, and is a consequence of the flaw in Arora et al.'s methodology.

\eat{
\note[Wei]{To-do: It would be nice to point the reader to some plots in TIM and IMM, where we show that there is no need to go to 0.05 for $\epsilon$, and at values such as $0.2$, the spread quality is still very good and the running time is much much faster.}

\note[Xiaokui]{Sure! The TIM and IMM papers do not have any plot for the IC model with uniform edge weight = 0.1, but Keke is working on some experiment to generate some plots.}
}

%\note[Amit]{@Xiaokui, should we highlight this misclaim as well? If yes, can you write something here? Thanks :-).}
%
%\note[Wei]{In the Simpath section, I labeled each misclaim using (SP1), (SP2), and put them in quote environment. I realize that (SP) may be too subtle.  To unify the format, We can just define a ``theorem'' like environment, with ``Misclaim'' as the environment heading.}

\begin{misclaim}
\textit{``TIM$^+$ is faster than IMM under LT model.''}
-- \cite{agr17}, Section 5.3.1, Section 6, M3 (repeated).
\end{misclaim}

\noindent\textbf{Refutation:}
Arora et al.\ compare the empirical running time of TIM$^+$ and IMM under the LT model by setting $\epsilon = 0.1$ for TIM$^+$ and $\epsilon = 0.05$ for IMM, and they conclude that TIM$^+$ is empirically more efficient than IMM under the LT model.
Unfortunately, the comparison is meaningless and the conclusion made is invalid, as we explain in the following.

Both TIM$^+$ and IMM rely on drawing subgraph samples (i.e., RR-sets~\cite{borgs14}) for influence maximization. For both algorithms, the number of samples used is decided based on the given {\em theoretical worst-case accuracy threshold $\epsilon$}, and the running time of the algorithms almost solely depends on the sample number. IMM offers a tighter asymptotic bound of sampling accuracy than TIM$^+$ does, and hence, for a given $\epsilon$, it can use a smaller sample set than TIM$^+$ to achieve a $(1-1/e-\epsilon)$-approximation.
%Note that as $\epsilon$ increases, error in estimating spread (aka, accuracy) decreases.
In other words, IMM is more efficient than TIM$^+$ when they are held to the same {\em worst-case accuracy threshold $\epsilon$}, as convincingly demonstrated in \cite{tang15}.

In contrast, Arora et al.\ compare the efficiency of TIM$^+$ and IMM under the LT model with $\epsilon = 0.1$ for TIM$^+$ and $\epsilon = 0.05$ for IMM\footnote{The root cause of selecting different $\epsilon$ values goes back to their ill-designed experimental methodology, which is highlighted in Section \ref{sec:ed}}, which is meaningless given the differences in the $\epsilon$ value, as different $\epsilon$ values represent different approximation guarantees of the solutions.
A slightly more sensible comparison is to evaluate the computation cost of the two algorithms when they achieve the same {\em empirical} accuracy. However, in that case, the comparison is {\em moot} and should lead to only one conclusion: they have roughly the same running time when their empirical accuracies are the same. The reason is that the {\em empirical} accuracy of the two algorithms depends only the sizes of the sample sets that they use. If the experiment is set up in such a way that the two algorithms will yield the same empirical accuracy, then the number of samples that they use should be roughly the same, in which case their running time would also be similar. Therefore, it is incorrect to claim that TIM$^+$ is more efficient than IMM under the LT model, regardless of whether the accuracy metric concerned is the empirical expected spread or the worst-case accuracy threshold $\epsilon$.

\begin{table*}[ht!]
\centering
\begin{tabular}{|c|c|c|c|c|c|c|c|c|c|c|c|c|c|}
\hline
$\epsilon$ & 0.05 & 0.1 & 0.15 & 0.2 & 0.25 & 0.3 & 0.35 & 0.4 & $\cdots$ \\
\hline
\textbf{Chebyshev} & 400000 &	100000 &	44445 &	25000 &	16000 &	11112 & 8164 & 6250 & $\cdots$ \\
\hline
\textbf{Chernoff} & 18243 &	4561 &	2027 &	1141 &	730 &	507 &	373 &	286 & $\cdots$ \\
\hline
\end{tabular}
\caption{Sample size required by Chebyshev's inequality and Chernoff's bound for various $\epsilon$ ($\mu = \sigma = 1/2, \delta = 10^{-3}$).}
\label{tbl:cheby-chernoff}
\end{table*}

\spara{Profound nature of Arora et al.'s mistakes}
In fact, comparing the empirical accuracies of TIM$^+$ and IMM is analogous to comparing the empirical accuracies of Chernoff's bound and Chebyshev's inequality when estimating the mean of a random variable.
Such a comparison is not useful at all,
%and to the best of our knowledge, mathematicians never make such comparisons,
since the empirical accuracy of the estimation would be the same as long as the sample set size is the same.
Recall that Chernoff's bound is tighter than Chebyshev's inequality, if one aims to achieve a given theoretical \emph{guarantee}.

%\note[Amit]{Above, should it be ``roughly'' same running, or ``exactly'' same running time. }
%%\note[Wei]{minor issue in style: TIM$^+$ or TIM$^+$?}
%
%\note[Xiaokui]{There is a subtlety here. TIM ``throws away'' some of the RR sets that it generates (since it relies on the Chernoff bound and requires RR sets to be totally independent), whereas IMM utilizes each and every RR set that it produces. Therefore, even for the same empirical accuracy, the number of RR sets required by IMM is slightly smaller (since it does not throw away RR sets). However, this issue is too technical to explain and may confuse the reader. Therefore, I used ``roughly'' to simply the discussion a bit.}
%
%
%\note[Amit]{Please check ``to the best of our knowledge, mathematicians never make such comparisons''. This is a strong claim - please verify it.}
%
%\note[Xiaokui]{Maybe we can remove it, just to be safe?}

%\pink{\bf (=== From here to end of this subsection, is the new part added by Xiaokui ===)}

Ironically, Arora et al.'s experimental design may even lead to an obviously incorrect conclusion that Chebyshev's inequality is superior to Chernoff's bound, as we explain in the following. Let $x_1, x_2, \ldots, x_n$ be i.i.d.\ samples drawn from
%a
%probabilistic
%standard normal distribution on
some distribution over $[0, 1]$ with mean $\mu$ and variance $\sigma^2$. Chebyshev's inequality says that for any $c > 0$,
$$\Pr\left[\left|\frac{1}{n}\sum_{i = 1}^n x_i - \mu\right| \ge c \right] \le \frac{\sigma^2}{n c^2},$$
which is equivalent to
$$\Pr\left[\left|\frac{1}{n}\sum_{i = 1}^n x_i - \mu\right| \ge \epsilon\mu \right] \le \frac{\sigma^2}{n \epsilon^2\mu^2}.$$

Meanwhile, Chernoff's bound says %that
$$\Pr\left[\left|\frac{1}{n}\sum_{i = 1}^n x_i - \mu\right| \ge \epsilon\mu \right] \le 2 \exp\left(-\frac{n \mu \epsilon^2}{3}\right).$$

In other words, from Chebyshev's inequality, if we are to use $\frac{1}{n} \sum_{i=1}^n x_i$ as an estimation of $\mu$ and we aim to achieve $\epsilon$ relative error with at least $1 - \delta$ probability, then the number $n$ of samples required should satisfy
$$n \ge \frac{\sigma^2}{\delta \cdot \epsilon^2 \mu^2}.$$

Meanwhile, if we apply Chernoff's bound instead, we have
$$n \ge \frac{3 \log(2/\delta)}{\epsilon^2 \mu}.$$
It can be verified that the number $n$ from Chernoff's bound is smaller whenever $3 \delta \mu \log(2/\delta) \le \sigma^2$. For example, Table~\ref{tbl:cheby-chernoff} shows the sample numbers required by Chebyshev's inequality and Chernoff's bound when $\mu = \sigma = 1/2$, $\delta = 10^{-3}$, and for various values of $\epsilon$.

Suppose that we are to compare the ``empirical efficiency'' of Chebyshev's inequality and Chernoff's bound, following Arora et al.'s methodology. In that case, we first fix $\delta$ (e.g., $\delta = 10^{-3}$) and, for Chebyshev's inequality (resp.\ Chernoff's bound), we set a threshold $\tau_1$ (resp.\ $\tau_2$) on the maximum absolute error allowed in the estimation of $\mu$. (This is in the same spirit of requiring an influence maximization algorithm to achieve an expected spread of at least $\mu^* - sd^*$.) For simplicity, assume that $\tau_1 = \tau_2 = \tau$.

Next, we vary $\epsilon$ in $\{0.05, 0.1, 0.15, 0.20, \ldots\}$. (Arora et al.\ use this set of $\epsilon$ values in their evaluation of TIM$^+$ and IMM.) For each $\epsilon$, we compute the number $n$ of samples required by Chebyshev's inequality (resp.\ Chernoff's bound), and measure the empirical error in the estimation of $\mu$ when we use $n$ samples from $\Omega$. Suppose that $\epsilon_1$ (resp.\ $\epsilon_2$) is the largest value of $\epsilon$ for which Chebyshev's inequality (resp.\ Chernoff's bound) has an empirical error no more than $\tau$. Then, we will compare the sample number $n_1$ required by Chebyshev's inequality with $\epsilon = \epsilon_1$ against the number $n_2$ required by Chernoff's bound with $\epsilon = \epsilon_2$, and we declare that Chebyshev's inequality is empirically more efficient if $n_1 < n_2$.

For the sake of argument, assume that when we use at least $10000$ samples from $\Omega$, the estimation of $\mu$ has at most $\tau$ error with at least $(1 - \delta)$ probability. Then, in the context of Table~\ref{tbl:cheby-chernoff}, we can see that $\epsilon_1 = 0.3$ and $\epsilon_2 = 0.05$. Accordingly, $n_1 = 11112$ and $n_2 = 18243$, which leads to the conclusion that Chebyshev's inequality is empirically ``more efficient'' than Chernoff's bound!

%% file: sec-simpath.tex
\begin{figure*}[ht!]
\centering
\begin{tabular}{cccc}
\hspace{-5mm}
\includegraphics[width=.24\textwidth]{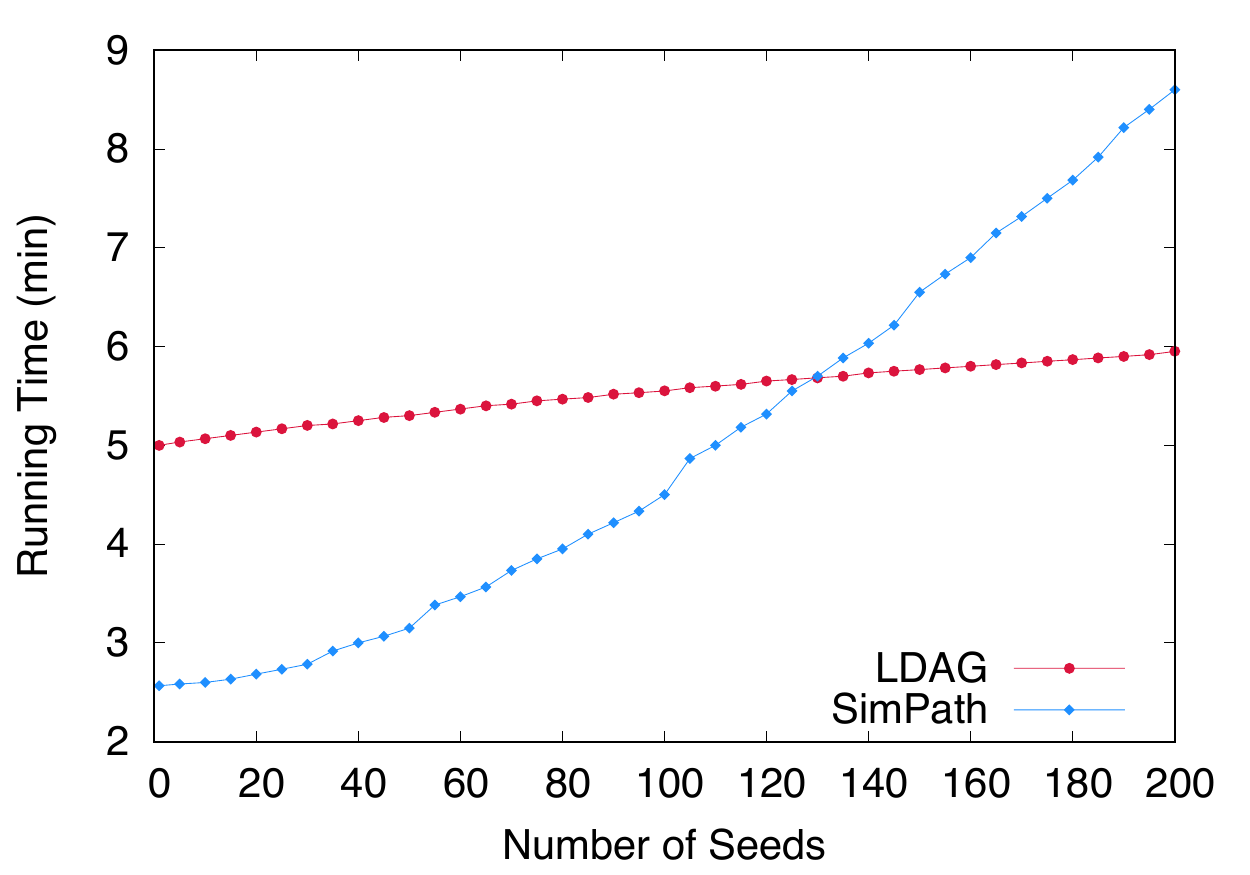}&
\hspace{-5mm}
\includegraphics[width=.24\textwidth]{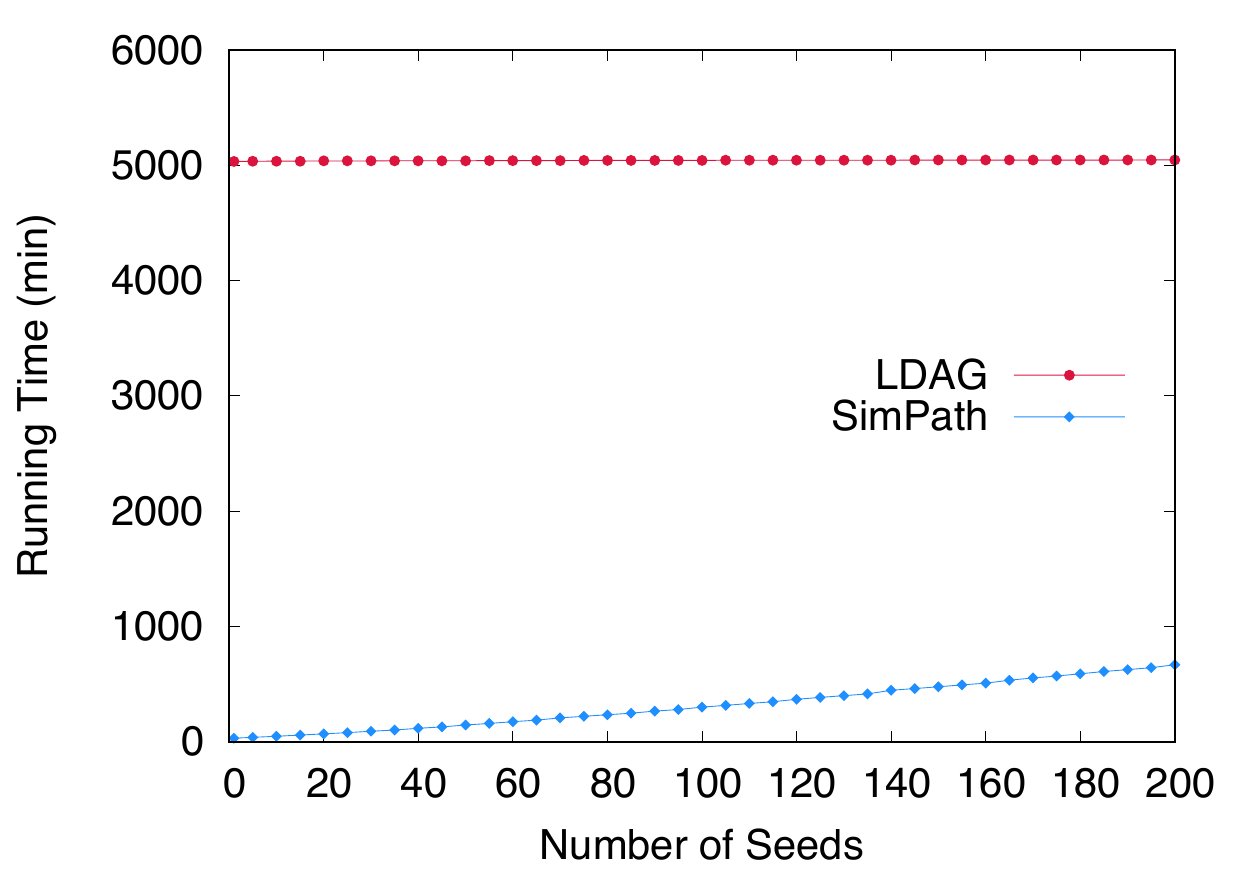} &
\hspace{-5mm}
\includegraphics[width=.24\textwidth]{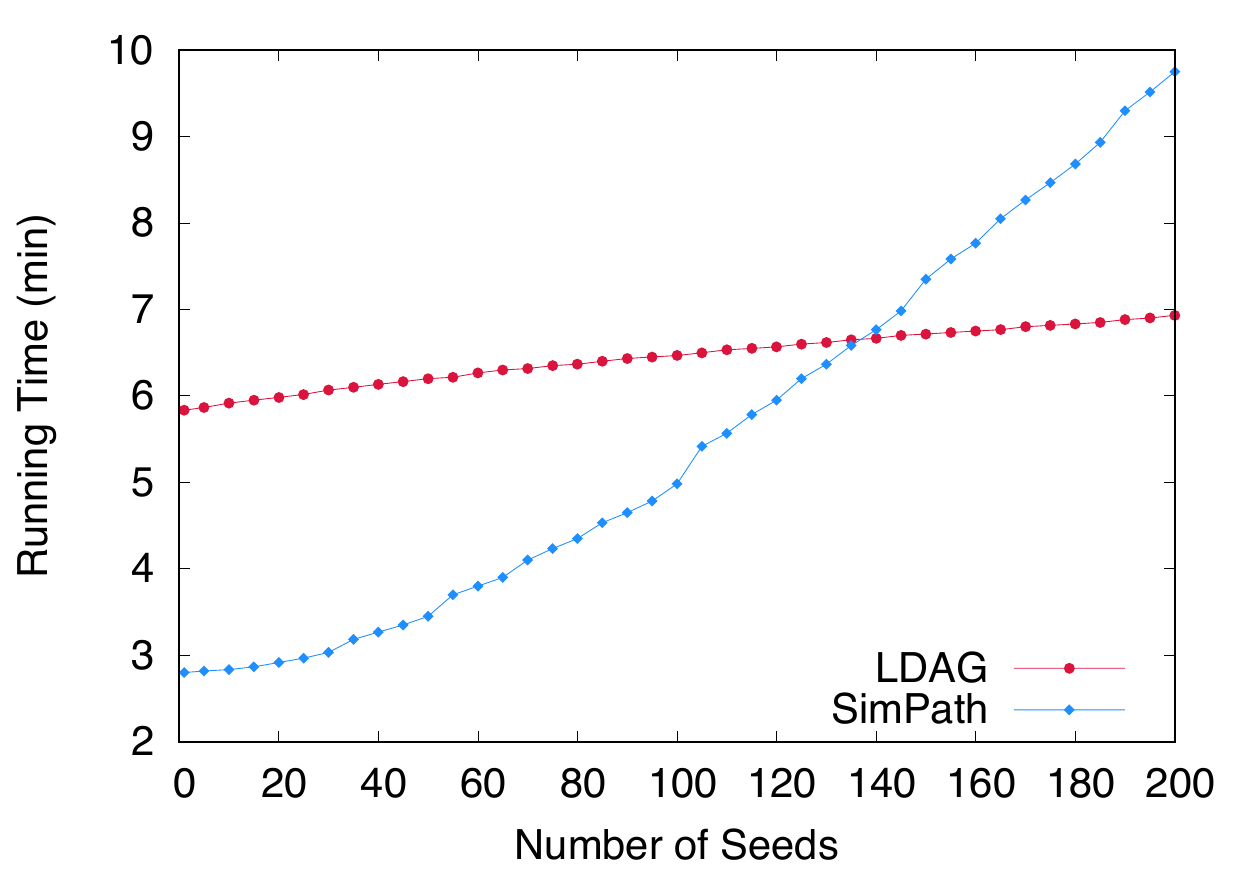} &
\hspace{-5mm}
\includegraphics[width=.24\textwidth]{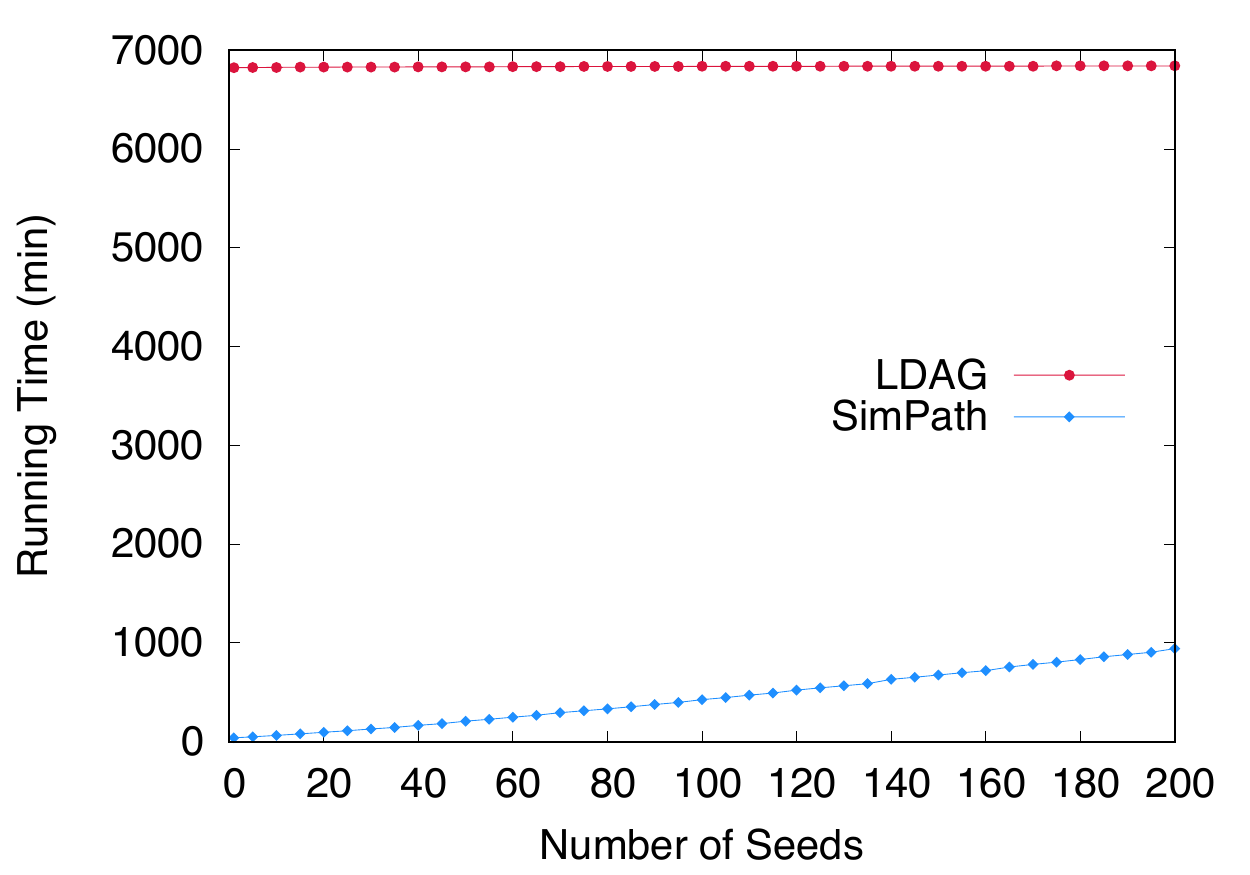}\\
(a). DBLP, UBC & (b). YouTube, UBC & (c). DBLP, NTU & (d). YouTube, NTU
 \end{tabular}
\caption{Running time of SimPath and LDAG, on DBLP and YouTube datasets. Note that the UBC server has faster CPUs and larger memory, and hence generally both algorithms run faster on the UBC server.}
\label{fig:ubc-simpath-time}
\end{figure*}

We now address the misclaims made by Arora et al.~\cite{agr17} against the SimPath algorithm~\cite{simpath}.
%Via efforts to attempt to reproduce their experiments, we have identified several incorrect claims.
Importantly, while attempting to reproduce their experiments, we obtained results directly contradict \cite{agr17}, and are in support of the original SimPath paper~\cite{simpath}.
We will explain the discrepancies.
We have verified all results presented in this subsection on both NTU and UBC servers\footnote{Unless otherwise stated, whenever running time numbers are mentioned in text, they are taken from the UBC results.}. 
%We shall also provide detailed reasons as to why significant discrepancies exist.
%Lastly, we give further insights regarding the comparisons of SIMPATH and LDAG, as observed from our new experiments.

\subsubsection{Misclaims in Arora et al. \& Refutations}\label{sec:simpath-claims}
%The following misclaims were presented in~\cite{agr17} regarding SimPath. % and the comparisons between SimPath and LDAG.
%We first state those claims in exact quotes and then give a succinct refutation to each of them.

\begin{misclaim}\label{mis:sp1}
\textit{``SIMPATH fails to finish even after 2400 hours on DBLP and YouTube.''}
-- \cite{agr17}, Section 5.3.1 and Section 6, M5 (repeated).
\end{misclaim}

\noindent\textbf{Refutation:} 
Unfortunately, Arora et al. failed to supply compatible datasets to the SimPath source code~\cite{ubc-code} released by Goyal et al.~\cite{simpath}, and ran into {\sl infinite loops} for at least 2400 hours. 
We have reached out to Arora et al., and they have acknowledged this issue. In particular, the code reserves node-id ``0'' for backtracking purposes, and therefore assumes that all input node-ids are positive integers. However, the datasets used by Arora et al.\ contain 0 as node-ids. They neither preprocessed the datasets correctly, nor did they attempt to debug it when their experiments were stuck in an infinite loop for 100 days.

With corrected input data, we verified that SimPath finishes within just a tiny fraction of 2400 hours: It took just 8.6 minutes on DBLP and 667 minutes on YouTube to select 200 seeds, representing 0.00597\% and 0.463\% of 2400 hours, respectively (Figure~\ref{fig:ubc-simpath-time})! 

%\vspace{2mm}  
\begin{misclaim}\label{mis:sp2}
\textit{``We discover that SIMPATH provides faster performance than LDAG only on the ``parallel edges'' LT model, which is used in the SIMPATH paper [15]. In our experiments, we use the ``uniform'' LT model.''} 
-- \cite{agr17}, Section 5.3.1 and Section 6, M5 (repeated).
\end{misclaim}

\noindent\textbf{Refutation:} Since the two datasets, DBLP and YouTube, on which Arora et al. ran into infinite loops happen to be prepared according to the so-called ``LT-uniform'' model, Misclaim \ref{mis:sp2} is in fact an corollary of Misclaim \ref{mis:sp1}, and thus incorrect.

%\vspace{2mm}
\begin{misclaim}\label{mis:sp3}
\textit{``In the SIMPATH paper, these two techniques are not evaluated beyond 100 seeds and thus this observation remained hidden. ''}
-- \cite{agr17}, Section 6, M5.
\end{misclaim}

\noindent\textbf{Refutation:} Almost all major IM work published prior to or shortly after SimPath~\cite{kempe03, KimuraS06, ChenWW10, ChenWW10b, ChenWY09, irie, timeinf} conducted experiments with $k=50$ seeds or less, including the LDAG paper itself~\cite{ChenWW10b}.
Hence, Misclaim \ref{mis:sp3} is ignorant of the historical context of the research in this domain. 
Moreover, the very trend that the running time of LDAG grows slower than SimPath as $k$ increases can be observed directly from the original SimPath paper ({\em cf.} Figure 4 in \cite{simpath}).
The usage of ``hidden'' by Arora et al. is misinformed and misleading.

%\vspace{2mm}
\begin{misclaim}\label{mis:sp4}
\textit{``Overall, these results indicate that LDAG not only scales better than SIMPATH but is also more robust to the underlying diffusion model.''}
-- \cite{agr17}, Section 6, M5.
\end{misclaim}
%\end{quote}

\noindent\textbf{Refutation:} 
Our results on the YouTube dataset directly refute this misclaim: Figure~\ref{fig:ubc-simpath-time} shows that when $k=200$, SimPath took 667 minutes while LDAG took 5047 minutes, which amounts to a gap of 7.5x favoring SimPath. 
As refuted in Misclaim \ref{mis:sp2}, the distinction between the so-called ``LT-uniform'' and ``LT-parallel'' models actually stems from an infinite loop, in turn the result of not preprocessing the data correctly. Therefore, LDAG being ``more robust to the underlying diffusion model'' is incorrect.
%, as it simply is a result of the infinite loops which Arora et al. ran into and remained unaware of for a very long time.
%Hence, it is a made-up distinction and is incorrect.

%\vspace{2mm}
\begin{misclaim}\label{mis:sp5}
\textit{``Note that LDAG [6], IRIE [16] and SIMPATH [15] do not have any external parameters, and thus this analysis does not cover them.''}
-- \cite{agr17}, Section 5.1.1.
\end{misclaim}

\noindent\textbf{Refutation:}
Arora et al. define the external parameter as follows\footnote{The definition of ``external parameters'' can be seen from the following exact quotes in \cite{agr17}: Section 3.1.1 states {\it ``Majority of IM algorithms M possess an external parameter which controls its accuracy... The more stringent the choice of this parameter the better the accuracy. Consequently, the more the running time.''} In addition, Section 5.1.1 states {\it ``The external parameters are exposed through the API and can be tuned to optimize performance.''}}:
It controls the quality and running time trade-off for the algorithm, and can be explicitly configured by the user.
%Misclaim \ref{mis:sp5} is obviously wrong -- 
Unfortunately, the authors have overlooked the fact that both LDAG and SimPath do have such a parameter, and both papers clearly stated so.
%\squishlisttight
\begin{itemize}[leftmargin=*]
\vspace{-1mm}
\item LDAG: Section IV.B, \cite{ChenWW10b}: ``... controlling the size of the
LDAG using parameter $\theta$, which represents a tradeoff between
efficiency (smaller DAGs and thus faster computations)
and accuracy (larger DAGs and more accurate influence
result).''
\vspace{-1mm}
\item SimPath: Section I.A, \cite{simpath}: ``We propose a parameter
$\eta$ to control the size of the neighborhood that represents a
direct trade-off between accuracy of spread estimation and running time.'' In fact, Table III in \cite{simpath} shows the trade-off between spread achieved and running time by varying parameter $\eta$, on two different datasets.
\vspace{-1mm}
\end{itemize}

In the released code \cite{ubc-code}, both $\theta$ in LDAG and $\eta$ in SimPath can be set or changed in a plain-text configuration file;
no code change is required.

\subsubsection{Reasons for Discrepancies}\label{sec:simpath-cause}

As mentioned, Misclaims~\ref{mis:sp1}, \ref{mis:sp2}, and \ref{mis:sp4} are caused by the fact that Arora et al.~\cite{agr17} did not use compatible datasets with our SimPath source code~\cite{ubc-code}, and got stuck in an  infinite loop for 100 days.
%The reason why a dataset could be incompatible is as follows.
To explain why an infinite loop may occur,
first recall that the SimPath algorithm has a backtracking procedure (Algorithm 2 in \cite{simpath}) to estimate expected influence spread.
The code makes use of integer ``0'' as a dummy node id for resetting the FIFO queue in backtracking.
If there exists a  node in the graph with real node id 0, then whenever backtracking is called on this node, an infinite loop will occur because the code cannot distinguish between the real node id 0 and the dummy id 0.

It is worth noting that depending on the graph structure, not all nodes may reach the backtracking stage. 
Therefore, the mere presence of node id 0 in the input graph may not always lead to infinite loops. 
Arora et al. apparently ran into infinite loops for two datasets: DBLP and YouTube. Unfortunately, they did not attempt to understand or debug the code and instead chose to continue running the experiments for 2400 hours (100 days)\footnote{We would like to gently remind the reader that any unsupported experimental code can encounter unforeseen issues, and users are encouraged to understand the code instead of applying it blindly. In case of unexpected results, debugging helps. Typically no warranty or author liability is attached to open source code, as in the case for our code as well.}.

%We would like to gently remind that any experimental codes can encounter unforeseen issues.
%To avoid situations like this in the future, we urge all users of our code that: 
%\begin{enumerate}
%\item  Understand the code first, and do not apply it blindly.
%\item  Make sure to read the terms and conditions before downloading.
%Typically no warranty or author liability is attached to open source code, and this is the case for our code as well.
%\end{enumerate}

\subsubsection{Experiments and Analysis}

We ran experiments on the DBLP and YouTube datasets (``LT-uniform'') to attempt to reproduce Misclaims \ref{mis:sp1}, \ref{mis:sp2} and \ref{mis:sp4}.
As mentioned earlier, we obtained results sharply contradicting the above misclaims.

The DBLP dataset contains 317K nodes and 1.05M undirected edges.
The YouTube dataset contains 1.13M nodes and 2.99M undirected edges.
See Table 1 in \cite{agr17} for detailed statistics. %, or the original source \cite{snap-stanford} .
In both datasets, the graphs are undirected.
As with \cite{agr17}, we direct all edges in both directions and for each resultant directed arc $(u,v)$, the influence weight $p_{u,v} = 1.0 / \text{indegree}(v)$ (as per the definition of ``LT-uniform'' in \cite{agr17}).
To avoid infinite loops, we preprocessed the data such that only positive integers are used as node ids.

We ran the experiments at both UBC and NTU, independently.
The executors used the same source code, configurations, and input graph files.
The UBC server runs OpenSuSE, has 16-core Intel Xeon X5570 CPUs at 2.93GHz each, and 94.4GB RAM.
The NTU server runs Debian, has 6-core Intel Xeon CPUs E5645 at 2.40GHz each, and 32GB RAM.
Every experiment (one particular algorithm on one particular dataset) was run as the only active process on the server, except for those required by the operating system.

\spara{Results and Analysis}
As with \cite{agr17}, we set $k=200$.
The running time obtained on the UBC and NTU servers are shown in Figure~\ref{fig:ubc-simpath-time}.
%Results obtained on the NTU server are shown in Figure~\ref{fig:ntu-simpath-time}.
Raw numbers can be viewed in our public GitHub repository\footnote{\url{https://github.com/jjboo/simpath-results}}.
The trends (growth of running time as $k$ increases) are consistent on both datasets.
The UBC server has faster CPUs and larger main memory, and hence the absolute values of the running time are smaller in Figure~\ref{fig:ubc-simpath-time}(a)(b) than Figure~\ref{fig:ubc-simpath-time}(c)(d).
To re-iterate, our results here directly and clearly refute Arora et al.'s misclaims on SimPath's running time.

%\begin{figure*}[ht!]
%\centering
%\begin{tabular}{cc}
%\includegraphics[width=.4\textwidth]{simpath-plots/ubc/dblp_wc_time.pdf}&
%\includegraphics[width=.4\textwidth]{simpath-plots/ubc/youtube_wc_time.pdf}\\
% (a) DBLP  & (b) YouTube
% \end{tabular}
%\caption{Running time of SimPath and LDAG, UBC server}
%\label{fig:ubc-simpath-time}
%\end{figure*}
%
%\begin{figure*}[ht!]
%\centering
%\begin{tabular}{cc}
%\includegraphics[width=.4\textwidth]{simpath-plots/ntu/dblp_wc_time.pdf}&
%\includegraphics[width=.4\textwidth]{simpath-plots/ntu/youtube_wc_time.pdf}\\
% (a) DBLP & (b) YouTube
%\end{tabular}
%\caption{Running time of SimPath and LDAG, NTU server}
%\label{fig:ntu-simpath-time}
%\end{figure*}

%Importantly, to re-iterate our refutations in Section~\ref{sec:simpath-claims}, 

\begin{itemize}[leftmargin=*]
\vspace{-2mm}
\item Refutation on Misclaim~\ref{mis:sp1}: %SimPath finish well before 2400 hours, it uses only a {\sl tiny fraction}:
SimPath took just 8.6 minutes on DBLP and 667 minutes on YouTube fo  selecting  200 seeds, representing 0.00597\% and 0.463\% of 2400 hours, respectively.
\vspace{-2mm}
\item Refutation on Misclaim~\ref{mis:sp2}: Both DBLP and YouTube datasets here are prepared under the ``LT-uniform'' definition. Figure  \ref{fig:ubc-simpath-time} shows that SimPath is clearly the winner on the larger YouTube data, taking 667 minutes to select 200 seeds, while LDAG takes 5047 minutes.
On DBLP data, SimPath is faster initially and LDAG catches up after approximately 130 seeds.
\vspace{-2mm}
\item Refutation on Misclaim~\ref{mis:sp4}: The 7.5x gap between SimPath's (667 minutes) and LDAG's (5047 minutes) running time on YouTube directly refutes Misclaim \ref{mis:sp4}. While LDAG is indeed a significant algorithmic contribution to IM under the LT model, it takes more than 3 days to finish selecting 200 seeds on a graph with 1.1M nodes whereas SimPath finished in 667 minutes.
\vspace{-2mm}
\end{itemize}

\spara{Further Remarks}
This exercise in fact illustrates different natures of SimPath and LDAG, and their respective advantages. LDAG does most of its work upfront, constructing a local directed acyclic graph for each node before going on to mine seeds. Once the local DAGs are built, the seed selection process is relatively fast. The SimPath algorithm, on the other hand, estimates the expected influence spread directly on the original input graph and intelligently enumerates and prunes simple paths to achieve this goal. As the seed set size increases, the number of paths that SimPath needs to examine increases, resulting in larger running time.
Given the very different nature of LDAG and SimPath, it is not surprising at all that on certain datasets, LDAG may regain advantages as the seed set size becomes larger.
%In fact, this very trend can be seen clearly in Figure 4 of \cite{simpath}, even though $k$ is capped at $50$ as a standard practice in IM research~\cite{kempe03, KimuraS06, ChenWW10, ChenWW10b, ChenWY09, irie, timeinf} {\em circa} 2011.

%To see the raw output files of the independent experiments done at both UBC and NTU, we refer the reader to our public GitHub repository at \url{https://github.com/jjboo/simpath-results}.

%\spara{Additional Experimental Results}
%We ran the experiments as shown in Figures \ref{fig:ubc-simpath-time} and \ref{fig:ntu-simpath-time} before Arora et al. provided all of their datasets files.
%The two graphs we prepared are isomorphic to theirs, but as a sanity check and a extra layer of verification, we also did the following exercise: adding 1 to each node id in their graphs, and rerunning SimPath.
%We obtained similar numbers on the UBC server, and shall report the numbers in Appendix~\ref{sec:app-simpath}.

%% file: sec-misc.tex
In addition to the list of serious issues discussed extensively in Sections~\ref{sec:ed} and \ref{sec:issues}, \cite{agr17} contains several  other incorrect, misleading, and/or unscientific statements.
Next, we shall give a {\em non-exhaustive} list and discuss them in detail.

\subsection{Misclaims about EaSyIM~\cite{easyim}}
EaSyIM is an algorithm proposed by Galhotra, Arora, and Roy in SIGMOD 2016~\cite{easyim}. Arora et al.\ \cite{agr17} make the following claims on the superiority of EaSyIM: 

\begin{misclaim}\label{mc:memory}
{\it ``EaSyIM [10] is most memory-efficient ... EaSyIM only stores a number per node. Consequently, it is the most memory efficient technique for IM.''} -- \cite{agr17}, Section 5.4.
\end{misclaim}

\begin{misclaim}\label{mc:dt}
``Fig. 11b presents the decision tree for choosing the best IM technique given the task and resources in hand...  When main memory is scarce, EaSyIM, CELF, CELF++ and IRIE provide alternative solutions. Among these, EaSyIM easily out-performs the other three techniques in memory footprint, while also generating reasonable quality and efficiency. Overall, the choice is between four techniques: IMM, TIM$^+$, EaSyIM, and PMC.'' -- \cite{agr17}, Section 7
\end{misclaim}

\spara{Refutation}
Arora et al. \cite{agr17} helpfully summarize their empirical findings in the form of a decision tree which is intended to make recommendations for what IM algorithm to use under what circumstances. Let us examine a couple of recommendations coming out of their decision tree and compare those recommendations with common sense.

Consider the IM problem under the LT model over a large dataset. Suppose available main memory is large. Then according to their decision tree, one should use TIM$^+$ and not IMM. We have already established in Section~\ref{sec:tim-imm} that under all major diffusion models including LT, IMM strictly dominates TIM$^+$ in the sense that for any desired \emph{theoretical worst case guarantee} $\epsilon$, IMM can deliver a seed set $S$ of size $k$ whose expected spread is guaranteed to be at least $(1-1/e-\epsilon)\cdot OPT$, w.h.p., with far fewer RR-sets than TIM$^+$ can. While TIM$^+$ can provide the same guarantee, it comes at the price of many more RR-sets than needed by IMM. If following Arora et al.'s argument we go ahead and use TIM$^+$ instead of IMM, it is true that \emph{sometimes} one may obtain the same \emph{empirical} accuracy as IMM with fewer RR-sets than dictated by the worst-case guarantee. But this accuracy is not always  guaranteed. Thus, this recommendation can lead to a poor choice of algorithm.

As a second example, in that decision tree, Arora et al.\ recommend EaSyIM~\cite{easyim} as the best choice for all three diffusion models (i.e., IC, WC, and LT) when ``memory is scarce'', and claim that EaSyIM is ``generating reasonable quality and efficiency'' without rigorously defining what exactly is meant by ``reasonable''. This recommendation is misleading and questionable for several reasons. 

First, Table 3 in \cite{agr17} itself presents contradicting results: For the WC and LT settings, EaSyIM ``did not terminate even after 40 hours''\footnote{Exact quote from the caption of Table 3 in \cite{agr17}.} on Orkut, Twitter, and Friendster. The only dataset in Table 3 that EaSyIM can handle is LiveJournal, which is only 2.5GB in size. This demonstrates that EaSyIM does not really offer reasonable efficiency or scalability, certainly not comparable to IMM or TIM$^+$ which finish on all datasets. Second, given that EaSyIM cannot handle the datasets larger than 2.5GB under WC and LT, it cannot really be recommended whenever ``memory is scarce''. For example, consider that we have a 4GB dataset and 5GB memory. The memory size is relatively small in comparison to the dataset size, and yet, EaSyIM would not be able to process the data due to its excessive computation cost. Third, given that nowadays, desktop computers (resp.\ workstations) can easily have more than 8GB (resp.\ 32GB) of memory, it is unrealistic to assume that one needs to process a smaller-than-2.5GB dataset with a machine whose memory is small with respect to the data.

Further, Misclaim~\ref{mc:memory} declares {\sl in absolute terms} that EaSyIM~\cite{easyim} is the most memory-efficient technique for IM. The rationale is that EaSyIM requires just a single number to be stored per node. 
So, by the same argument, one could prefer the Random algorithm, which randomly chooses $k$ seeds, over \emph{all} algorithms, including EasyIM, since Random does not require any information to be stored per node.
At another extreme, basic Greedy with MC simulations can also be chosen over EasyIM, as it also stores only one number per node.
Consider the extreme nature of these choices. Random is extremely fast but can lead to poor spread. Greedy with MC simulations can lead to very high spread depending on the number of simulations employed, but is extremely slow. Thus, an argument for choosing an algorithm solely based on its low memory consumption is ill-conceived as it ignores other equally important factors like running time and spread.

\subsection{Misleading ``Myths'' and Other Issues}

\begin{misclaim}
\textit{``M2. CELF (or CELF++) is the gold standard for quality''}
-- \cite{agr17}, Section 6, M2.
\end{misclaim}

\noindent\textbf{Refutation.}
This is a myth Arora et al.~\cite{agr17} claimed to have found from the influence maximization literature.
Their argument is that the solution quality of CELF (or CELF++) depends on the number of Monte Carlo (MC) simulations.
First of all, CELF and CELF++ are both heuristic optimizations on top of the vanilla greedy approximation algorithm, that save on the number of marginal gain computations and have nothing to do with ``quality''.
It is well known that the quality solely depends on the number of MC simulations, which are used to compute the spread function, and which are orthogonal to the CELF and CELF++ optimizations. 
Hence, CELF or CELF++, per se, has nothing to do with quality.
In particular, it is strange to suggest that a heuristic, for saving on marginal gain computations, in and of itself can be a standard of anything leave alone gold standard of quality!

\begin{misclaim}
{\it``M6. WC is equivalent to IC. Several techniques have misused the term IC ... they claimed to be the state-of-the-art for IC. In reality, they all fare poorly on the generic IC model ...''}
-- \cite{agr17}, Section 6, M6.
\end{misclaim}

\noindent\textbf{Refutation.}
It is correct that IC and WC indeed have different meanings.
However, the comments of Arora et al., quoted above, are misinformed and misleading.

IC refers to the ``Independent Cascade'' model, while ``WC'' stands for ``Weighted Cascade'', which is one method to compute influence probabilities under the IC model, i.e., $p_{u,v} = 1/\mathrm{indegree}(v)$.
One can use the WC method to compute influence probabilities, and find seeds in the IC model. 
That said, there does exist an early work~\cite{ChenWY09} (2009) that freely used ``IC'' to refer to the two things collectively. Consider the following two methods of assigning influence probabilities: (1) the WC method; (2) assigning an identical influence probability of $0.1$ to all edges in the graph. Both (1) and (2) are just two different methods of assigning probabilities. Neither of them is generic. In particular, method (2) grossly oversimplifies reality as there is no reason to believe that all edges will have the same influence probability! 
\eat{
(1). The underlying diffusion model is Independent Cascade; (2). All edges in the graph have {\sl identical} influence probabilities, say 0.1, {\sl an arbitrary number which grossly over-simplifies} the input graph and in no way represents real-word scenarios. Notice that 
}
Surprisingly, Arora et al. refer to (2) {``\em general IC model''} and go on to use this as a tool to criticize many prior papers.

%%%%%%%%%%%%%%%%%%%%%%%%

\medskip
Finally, regarding the experimental platform used in \cite{agr17}, Arora et al. state that in Section 5 (emphasis added):
{\it ``All experiments are performed using codes written in C++ on an Intel(R) Xeon(R) E5-2698 64-core machine with 2.3 GHz CPU and 256 GB RAM running {\tt Ubuntu 14.04}.'' and
``IRIE [16] was compiled on a {\tt Microsoft Windows 7} machine possessing the same configuration.''}

\spara{Observations \& Remarks}
Essentially, Arora et al.~\cite{agr17} ran one algorithm (IRIE) on Windows 7 while all other algorithms on Ubuntu (Linux).
Then the running time and memory usage of IRIE are squarely compared together with all other Linux-run algorithms (see Figure 7 and 8 of \cite{agr17}).
This is clearly an unscientific approach to benchmarking.
At the very least, a benchmarking experiment should be done using the {\sl same} operating system platform.
This is especially critical for C++ as the compilation and optimization of C++ code are vastly different for Linux (e.g., gcc) and Windows (e.g., Microsoft Visual Studio).

%% file: sec-concl.tex
The benchmarking paper by Arora et al.~\cite{agr17} claims to unearth and debunk so-called ``myths'' that allegedly propagated in the Influence Maximization research literature over the years. However, our refutations show that not only their experimental methodology is ill-designed and flawed, many of their specific experimental results and claims are incorrect.
In particular, they fail to incorporate the trade-off between running time and the seed set quality correctly in their experimental methodology.
In addition, they fail to distinguish between theoretical guarantee on spread from empirical spread, resulting in misleading and often wrong recommendations of IM algorithms to use under given resource constraints.  

In this paper, we have systematically refuted the experimental methodology of Arora et al. and 11 of the specific claims made in their paper. 
%Further details can be found in the appendices. 

%% file: sec-prelim.tex
%\subsection{Background and Preliminaries}

%Domingos and Richardson first introduced viral marketing as a computational problem to the data mining field~\cite{domingos01, richardson02}.
%Kempe et al.~\cite{kempe03} formulated {Influence Maximization (IM)} as a discrete optimization problem, as we mentioned in Section~\ref{sec:intro}.

Two classical stochastic propagation models were studied by Kempe et al.\ in \cite{kempe03}: {\em Independent Cascade (IC)} and {\em Linear Thresholds (LT)}, both of which originally stem from mathematical sociology.
For the exact definitions of IC and LT models, we refer the reader to \cite{kempe03, laksbook}.
Kempe et al.~\cite{kempe03} showed that the problem of Influence Maximization (IM) is NP-hard under both models.
Further, Chen et al.~\cite{ChenWW10, ChenWW10b} showed that under both models, it is \#P-hard to compute the exact value of the expected influence spread $\sigma(S)$, for any seed set $S$.
As a result, IM is a computationally challenging problem.
Kempe et al.~\cite{kempe03} established that the spread function $\sigma(\cdot)$ is {\em monotone}\footnote{A set function $f$ is {\em monotone} if $f(S)\leq f(T)$ whenever $S\subseteq T$.} and {\em submodular}\footnote{A set function $f$ is {\em submodular} if $f(S\cup\{x\})-f(S) \geq f(T\cup\{x\})-f(T)$, for all $S$ and $T$ where $S\subseteq T$ and all $x\not\in T$.} under both IC and LT models.
Hence, by applying a seminal result in Nemhauser et al.~\cite{submodular}, one can use a simple greedy hill-climbing style algorithm to achieve an approximation factor of $(1-1/e-\epsilon)$, for any $\epsilon > 0$.
The existence of $\epsilon$ is because it is \#P-hard to compute the spread function $\sigma(\cdot)$ exactly, and hence Monte Carlo (MC) simulations need to be used jointly with the greedy algorithm.
The greedy approximation algorithm starts with $S = \emptyset$, and runs for $k$ iterations.
In each iteration, it selects into $S$ a node that yields the largest marginal gain, defined as $\sigma(S\cup\{u\})-\sigma(S)$. We refer to this as the simple {\sl Greedy algorithm with MC simulations}. 

%\begin{algorithm}[t]
%\DontPrintSemicolon
%\caption{Greedy approximation algorithm for IM} \label{algo:greedy}
%\KwData{graph $G=(V,E,p)$, seed set size $k$}
%\KwResult{seed set $S$ such that $|S| \leq k$}
%\Begin{
%    $S \gets \emptyset$\;
%    \For {$i \gets 1 \text{ to } k$} {
%    	$w \gets \argmax_{u\in V\setminus S} \left[ \sigma(S\cup\{u\}) - \sigma(S) \right]$\;  \label{line:greedyMG}
%    	$S \gets S \cup \{w\}$\;
%    }
%}
%\end{algorithm}	

%Kempe et al.~\cite{kempe03} used the greedy algorithm with MC simulations. % in their empirical studies.

The computational costs associated with MC simulations are inevitably high, rendering the greedy algorithm quite inefficient and unable scale to large graphs. Several notable improvements have been made since then \cite{leskovec07, ChenWW10, ChenWW10b, simpath, irie, laksbook, borgs14, tang14, tang15, jpei16, gomez13, gomez16, chengSHZC13, chengSHCC14, ohsaka16}, some of which are covered in this paper. We next recall them briefly.

%To set the stage for refutations on false claims by Arora et al.~\cite{agr17} regarding multiple IM algorithms, we first recap the technical aspects of each algorithm covered in subsequent sections of this work.
%The following list is purely in chronological order of publication.

%\subsection{Algorithms Covered}
%The following list is in chronological order of publication.

%\begin{description}
%\item [CELF~\cite{leskovec07}.]
\spara{CELF~\cite{leskovec07}}
The vanilla greedy algorithm coupled with MC simulations, is rather simplistic.
In each iteration, for all $w\in V\setminus S$ (all nodes in the graph except for those already selected as seeds), a re-computation of its marginal gain $\sigma(S\cup\{w\}) - \sigma(S)$ is done. % carried out.
%By marginal gain, we mean the incremental value a node offers to the spread function given the current seed set $S$.
CELF, for Cost-Effective Lazy-Forward, employs a clever optimization technique: It sorts nodes in non-increasing order of their latest marginal gain value (in a max-heap), and only recomputes the marginal gain of a node when this particular node surfaces to the root of the max-heap.
%By definition of submodularity, {\sl in theory}, given the same input, the greedy algorithm produces identical seed sets with and without CELF\footnote{In practice, however, the output may subject to variance and noise beyond human control. We shall elaborate in details in Section~\ref{sec:celfpp}.}.

%\item [CELF++~\cite{celfpp}]
\spara{CELF++~\cite{celfpp}}
A poster paper by Goyal et al.~\cite{celfpp} built on top of CELF and tried to further improve the greedy algorithm by leveraging submodularity ``more aggressively''.
Whenever an evaluation of marginal gain is carried out, CELF++ also does a speculative, look-ahead marginal gain computation.
More specifically, for each node $w$, CELF++ records the current best node in the max-heap (let us call it $w^*$) and computes $\sigma(S\cup\{w\}) - \sigma(S)$ and $\sigma(S\cup\{w, w^*\}) - \sigma(S\cup \{w^*\})$ together.
Then, by definition of submodularity, as long as $w^*$ indeed gets chosen as a seed, $\sigma(S\cup\{w, w^*\}) - \sigma(S\cup \{w^*\})$ will be readily available. {\sl Notice that both CELF and CELF++ are heuristics that help save on the number of marginal gain computations and are completely orthogonal to the MC simulations. It is obvious that the quality of expected spread estimation is controlled by the \# MC simulations used and has nothing to do with CELF or CELF++.}  

%\item [LDAG~\cite{ChenWW10b}]
\spara{LDAG~\cite{ChenWW10b}}
Chen et al.~\cite{ChenWW10} proposed the LDAG (stands for {\em Local Directed Acyclic Graph}) heuristic for solving IM under the Linear Threshold (LT) model.
The main intuitions is that (1).\ even though computing exact influence spread is \#P-hard in the LT model, it can be done in linear time on DAGs;
(2).\ as a heuristic, the influence propagates to and from a node can be estimated using a local neighborhood surrounding the node, instead of the whole graph.
Hence, a two-phase heuristic algorithm, LDAG, was proposed.
First, for each node $v\in V$, one local DAG structure is constructed.
Second, seeds are selected in greedy order with marginal gains estimated using the DAGs.
LDAG achieved impressive results -- it was shown to be orders of magnitude faster than the greedy algorithm (with CELF), without sacrificing much solution quality.

\spara{SimPath~\cite{simpath}}
LDAG has a few limitations, e.g., memory consumption and the slight aggressive nature of keeping only one DAG per node and allowing influence to only flow within that lone DAG.
Goyal et al.~\cite{simpath} proposed a new heuristic called SimPath, establishing that under the LT model, influence spread can be computed by enumerating simple paths starting from the seeds.
SimPath employs several heuristical optimizations to prune and cut off path enumerations, maintaining a balance between running time and seed set quality.
In their experiments, Goyal et al.~\cite{simpath} showed that SimPath outperforms LDAG in three metrics: spread achieved (quality), running time, and memory consumption.

\spara{TIM and TIM$^+$~\cite{tang14}}
The latest break-through in IM (approximation) algorithm design started from the notion of Reverse-Reachable sets (or, RR-sets in short)\footnote{Consider a deterministic graph $G=(V,E)$ (i.e., edge presence is binary instead of being probabilistic), and any arbitrary node $v\in V$. An RR-set, rooted at $v$, is the set of all nodes that can reach $v$.} by Borgs et al.~\cite{borgs14}.
%They
%Prior to 2014, almost all work on addressing the efficiency and scalability of IM resort to proposing heuristics, which come with no worst-case approximation guarantees.
%defined the notion of Reverse Reachable sets, or RR-sets in the IM context.
Utilizing RR-sets, Borgs et al.\ developed an approximate algorithm with near-optimal time complexity, but it incurs significant computation overheads in practice.

Building upon Borgs et al.'s work, Tang et al.~\cite{tang14} designed the Two-phase Influence Maximization (TIM) algorithm that improves Borgs et al.'s solution in terms of both time complexity and practical efficiency. Given any IM instance, the first phase computes a lower-bound on the optimal spread, which is in turn used to determine the number of RR-sets, $\theta$, that it should sample.
The second phase, it samples $\theta$ RR-sets and returns a cardinality-$k$ node-set which covers the most sampled RR-sets, as the seed set. For both IC and LT models produces, TIM provides $(1-1/e-\epsilon)$-approximate solutions with at least $1-|V|^{-\ell}$ probability.

TIM$^+$ is an improved version of TIM. It has an intermediate step to refine the estimation of the lower bound on the optimal spread, and hence yields a smaller $\theta$ (which means fewer samples and better efficiency).
Both TIM and TIM$^+$ have expected time complexity $\mathrm{O}((k+\ell)(|E|+|V|)\log|V|\epsilon^{-2})$ and are orders of magnitude faster than the greedy algorithm.
They achieve even better efficiency than fast heuristics such as SimPath~\cite{simpath} and IRIE~\cite{irie}.
%The theoretical guarantees and time complexity of TIM/TIM+ have been stated in Section~\ref{sec:prelim}.

\spara{IMM~\cite{tang15}} IMM is an improvement of TIM$^+$ that offers significantly higher efficiency in practice while retaining the latter's asymptotic guarantees (i.e., IMM also returns $(1-1/e-\epsilon)$-approximations in $\mathrm{O}((k+\ell)(|E|+|V|)\log|V|\epsilon^{-2})$ expected time). Its main difference from TIM$^+$ is that it adopts a more advanced martingale-based approach to derive a tighter lower bound on the optimal spread, which enables it to achieve the desired approximation guarantee with a smaller number of RR sets than TIM$^+$. Tang et al.\ show that IMM is up to 100 times faster than TIM$^+$ when they achieve the same asymptotic assurance.

%% file: sec-celfpp.tex
%\note[Amit]{Explain CELF and CELF++ a bit.}
%\note[Wei]{Amit, no need as it is done in appendix A}

Arora et al.\ \cite{agr17} in their experiments observed that there are no significant differences in running times of CELF and CELF++, contradicting the observations in the original CELF++ poster paper \cite{celfpp}. 
To verify their claims, % from Arora et al.~\cite{agr17}
we reimplemented both CELF and CELF++ algorithms.
%\footnote{All new code is available on \url{https://github.com/jjboo/InfMax}} and reran the experiments that were presented in the original CELF++ paper \cite{www_poster}. 
In particular, we observed that there exists non-trivial variance in the running times of both CELF and CELF++.
After careful analysis, we observed that several factors contribute to this variance: $(i)$ Many nodes tend to have very close spread or marginal gains (on the two settings we tried). Different runs of the same algorithm can make different choices while selecting seeds, making the rest of the seed selection computations very different. $(ii)$ Different initial starting values for random number generation can lead to different choices, and thereby different running times. $(iii)$ The running time varies significantly if multiple experiments are run at the same time, even on a multi-core server, as in many CPU designs, the L2 and L3 caches may be shared across all cores.
%\footnote{\url{https://en.wikipedia.org/wiki/Smart_Cache}}.

Therefore, in order to obtain accurate running time and a statistically significant result, it is required that only one experiment is run at a time, even if the machine has multiple cores and each experiment is run multiple times to get a distribution of running times. A statistical significance test can then be applied to determine if the different in running time is statistically significant.

\begin{figure}[t!]
\centering
\begin{tabular}{cc}
\hspace{-2mm}
\includegraphics[width=.24\textwidth]{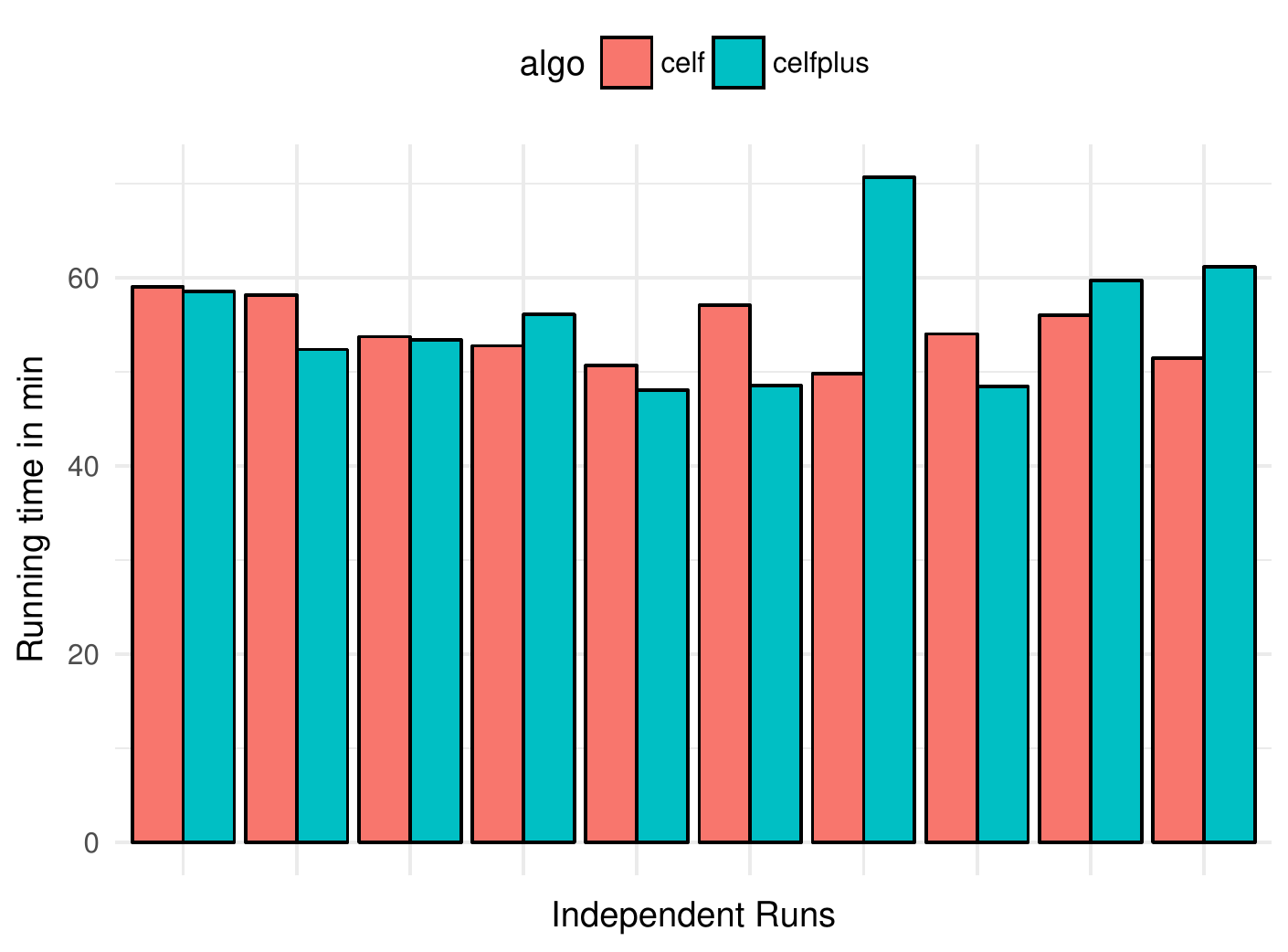}&
\hspace{-2mm}
\includegraphics[width=.24\textwidth]{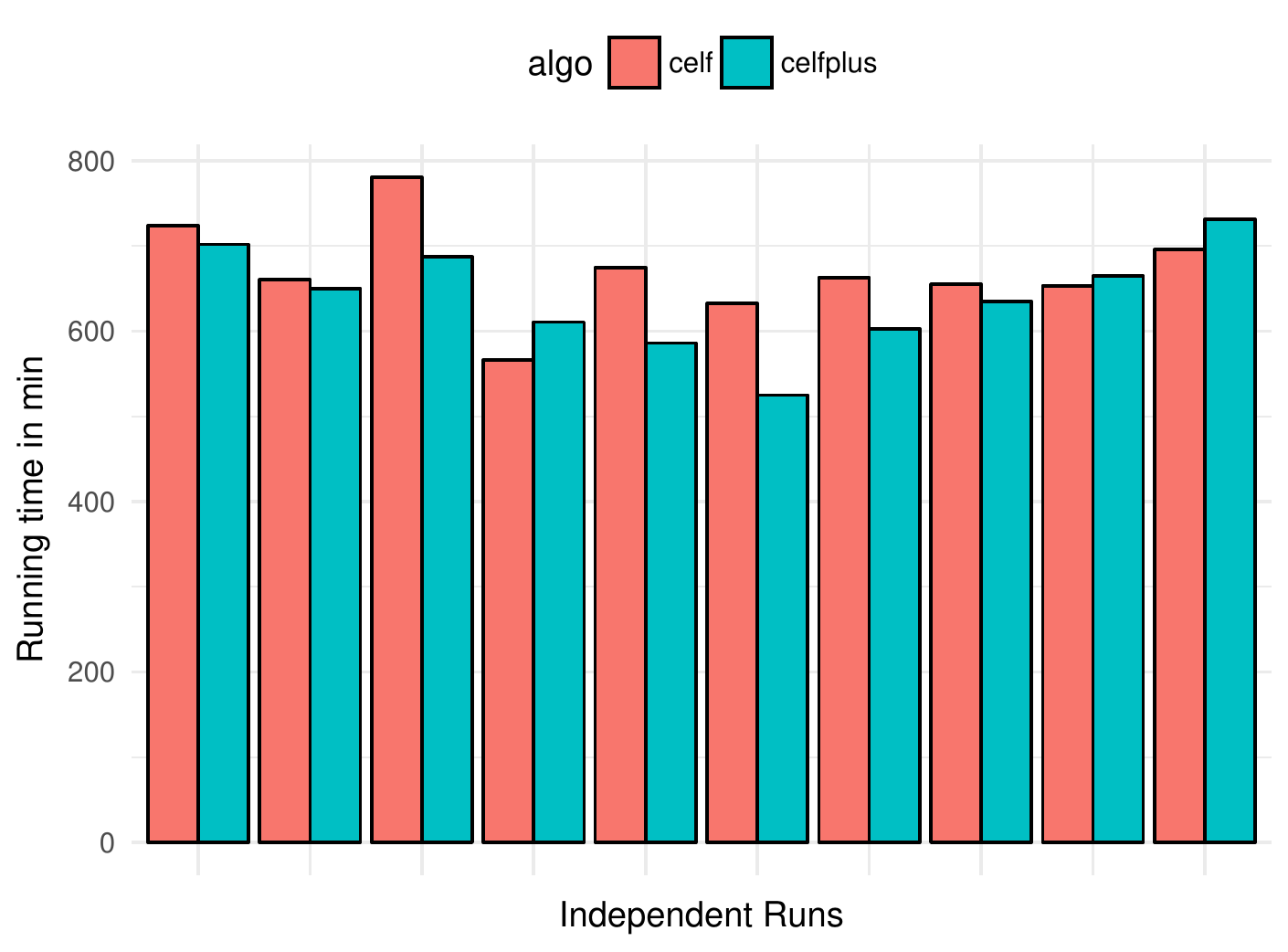}\\
 (a) NetHEPT WC & (b) NetHEPT IC
 \end{tabular}
\caption{Running time CELF and CELF++ on NetHEPT, 10 independent \& isolated runs each}
\label{fig:celfpp}
\end{figure}

In the experiments we report here, to avoid these issues, we ran exactly one experiment at a time on a machine\footnote{In fact, all experiments reported in this paper are run in isolation: only one experiment is run on a machine at a time.}. In addition, to maintain fairness in comparisons, we randomly selected 10 starting values for random number generation that were used both in CELF and CELF++ in their 10 respective independent runs\footnote{All code, including the algorithm to select starting values of random number generation is available at our GitHub repository: \url{https://github.com/jjboo/InfMax}. The MIT license applies.}. 
The results are shown in Figure \ref{fig:celfpp}. From 10 runs, under NetHEPT WC settings, the average running times of CELF and CELF++ are almost identical - 54.2 min and 55.7 min respectively. The $p$-value from $t$-test is 0.58 confirming that the difference is not statistically significant. 
Recall that a lower $p$-value implies higher significance in difference. Commonly, a $p$-value of 0.05 or lower is considered suitable for establishing the difference to be statistically significant. In case of NetHEPT IC, with influence probabilities 0.1 on all edges, average CELF running time is 670.5 minutes while it is 639.5 minutes for CELF++.
Here, CELF++ is faster than CELF by 4.6\%, but the $p$-value is 0.25, making it not statistically significant either. 

%It is worth mentioning that it may require around 25-30 runs (instead of 10) to establish statistical significance in 4.8\% difference with similar variance.

%It is worth mentioning that to obtain a statistical significance result for a difference of 5\%..

%\note{Include the two plots and state their $p$-values.}

%CELF running time mean WC = 54.3 min. 
%CELF++ running time mean WC = 55.7 min. 
%$p$-value = 0.58

%CELF running time mean IC = 670.5 min. 
%CELF++ running time mean IC = 639.5 min. 
%$p$-value = 0.25
We hereby acknowledge that results reported in CELF++ poster paper \cite{celfpp} ran into noise and are not statistically significant. We would like to thank Arora et al.~\cite{agr17} for bringing up the issue to our attention. 
%Currently, we are developing suitable statistical significance tests, and shall report updated results as soon as practicable.

%% file: sec-appendix-unscientific.tex
Two exact quotes from  Arora et al.:

\begin{itemize}
\item {\it (a) ``In this study, we need to either gather code from the authors or re-implement them. '' -- \cite{agr17}, Section 1}

\item {\it (b) ``Furthermore, to integrate them into the benchmarking framework and interpret the results, it is critical to have an in-depth understanding of the code.'' -- \cite{agr17}, Section 1}
\end{itemize}

\spara{Our Observations}
Even though statement (a) is not logically incorrect per se (due to the use of ``or''), it is not clear from \cite{agr17} which algorithm that Arora et al.\ had to ``re-implement''.
The implementation of EaSyIM~\cite{easyim} may come from Arora or Galhotra, but for the purpose of this paper, no re-implementation is needed.
From Arora et al.'s shared repo\footnote{Shared by Arora et al: \url{https://drive.google.com/drive/u/1/folders/0B3hfiezv112RUEVVUlpraGRQNWc}. Last accessed on May 12, 2017. A zipped copy of all files (version as of May 12, 2017) contained in the linked Google Drive folder is available upon request should this link become inactive.}, all other algorithms' implementations were made available to them by other research groups \cite{simpath, ChenWW10b, irie, celfpp, tang15, tang14, chengSHZC13, chengSHCC14,ohsaka16}.

Regarding statement (b), unfortunately it appears that while Arora et al.\ highlighted the importance of having an in-depth understanding of the code, they themselves did not adhere to it. In particular, as described above, a couple of their SimPath experiments were stuck in infinite loop for 100 days, and yet, they did not attempt to understand the code or prepare the dataset properly during that whole time period.